\documentclass[12pt]{article}
\usepackage{amsmath,amssymb,bm,epsfig}
\usepackage{dcolumn}
\usepackage{longtable}
\newcolumntype{d}[1]{D{.}{.}{#1}}

\renewcommand{\thefootnote}{\fnsymbol{footnote}}

\begin{document}

\title{
\begin{flushright}
\begin{minipage}{0.2\linewidth}
\normalsize
WU-HEP-18-01 \\*[50pt]
\end{minipage}
\end{flushright}
{\Large \bf 
$SO(32)$ heterotic line bundle models
\\*[20pt]}}

\author{Hajime~Otsuka$^{1,}$\footnote{
E-mail address: h.otsuka@aoni.waseda.jp
}\\*[20pt]
$^1${\it \normalsize 
Department of Physics, Waseda University, 
Tokyo 169-8555, Japan} \\*[50pt]}

\date{
\centerline{\small \bf Abstract}
\begin{minipage}{0.9\linewidth}
\medskip 
\medskip 
\small
We search for the three-generation standard-like and/or Pati-Salam models from 
the $SO(32)$ heterotic string theory on smooth, quotient complete intersection Calabi-Yau threefolds 
with multiple line bundles, each with structure group $U(1)$.  
%The stable line bundles lead to the three chiral generations of quarks and leptons 
%without chiral exotics. The Higgs doublets appear as the vector-like particles 
%under the standard model gauge groups. 
%We aim to directly derive the standard-like and/or Pati-Salam models which are related to 
%the S- and T-dual models of intersecting D-branes in type IIA string theory. 
%These models are related to the S- and T-dual models of intersecting D-branes in type IIA string theory. 
These models are S- and T-dual to intersecting D-brane models in type IIA string theory. 
We find that the stable line bundles and Wilson lines lead to the standard model gauge group with an extra $U(1)_{B-L}$ 
via a Pati-Salam-like symmetry and the obtained spectrum consists of three chiral generations of quarks and leptons, and vector-like particles. 
Green-Schwarz anomalous U(1) symmetries control not only the Yukawa couplings of the quarks and leptons 
but also the higher-dimensional operators causing the proton decay. 
%We also discuss the allowed Yukawa couplings of quarks and leptons. 
%Proton decay operators are forbidden with the help of $U(1)_{B-L}$ and anomalous $U(1)$ symmetries. 
%The stable line bundles lead to the standard model and $U(1)_{B-L}$ gauge groups 
%via a Pati-Salam-like symmetry and the obtained spectra consist of three chiral generations of quarks and leptons, and vector-like particles. 
%It is found that the Yukawa couplings of quarks and leptons are allowed at the renormalizable level and 
%proton decay operators are forbidden with the help of $U(1)_{B-L}$ and anomalous $U(1)$ symmetries. 
\end{minipage}
}

\begin{titlepage}
\maketitle
\thispagestyle{empty}
%\clearpage
%\thispagestyle{empty}
\end{titlepage}

\renewcommand{\thefootnote}{\arabic{footnote}}
\setcounter{footnote}{0}
%\vspace{35pt}

\tableofcontents

%%%%%%%%%%%%%%%%%%%%%%%%%%%%%%%%%%%%%%%%%%%%%%%%%%%%%%%%%%%%%%%%%%%%%%%%%%%%%%%%%%%%%%%%%%%%%%%%%%%%%%%%%%%%%
\section{Introduction} 
String theory is a most successful candidate of a unified theory including 
both the gauge and gravitational interactions. 
Among the perturbative superstring theories, the heterotic string theory~\cite{Gross:1985fr,Gross:1985rr} is an attractive one, 
since the gauge group and matter representations are 
uniquely determined by the ten-dimensional gauge and gravitational anomaly cancellation conditions. 
In the low-energy effective action of the heterotic string, 
the background gauge field strength is related to the curvature of internal manifold through the Bianchi identity. 
The simplest approach to solve the Bianchi identity is the ``standard embedding'', where the gauge bundle with $SU(3)$ structure group is identified with the holomorphic tangent bundle of the Calabi-Yau (CY) manifolds~\cite{Candelas:1985en}. (See, 
e.g. Refs.~\cite{Greene:1986bm,Greene:1986jb} for the detailed analysis of three-generation models.) 
On the other hand, it is possible to construct consistent heterotic models on smooth CY threefolds, 
even if the gauge bundles are not directly related to the tangent bundle~\cite{Witten:1985bz}. 
In this approach, holomorphic gauge bundles satisfying the Hermitian Yang-Mills equations lead to semi-realistic 
standard-like models as discussed in several CY threefolds with stable non-abelian vector bundles~\cite{Donagi:2000zf,Andreas:1999ty,Braun:2005ux,Blumenhagen:2005ga,Blumenhagen:2005pm,Bouchard:2005ag} and line bundles~\cite{Anderson:2011ns,Anderson:2012yf}.

So far, $E_8\times E_8$ heterotic string models 
have been well discussed in contrast to $SO(32)$ heterotic ones. 
This is because the adjoint representation of $E_8$ gauge group naturally involves the spectrum of 
$E_6$, $SO(10)$ and $SU(5)$ grand unified theories (GUTs). 
%(asymmetric) toroidal orbifold~\cite{Narain:1986qm},  Calabi-Yau (CY) threefold with line bundles~\cite{Anderson:2011ns} and 
%elliptically fibered CY threefold with stable vector bundles~\cite{Donagi:2000zf}.
%\footnote{See for the recent approach in non-supersymmetric heterotic string theory, e.g., Refs.~\cite{}.} 
%Since the adjoint representation of $E_8$ gauge group naturally involves the spectrum of grand unified theories 
%such as $E_6$, $SO(10)$ and $SU(5)$, 
On the other hand, the adjoint representation of $SO(32)$ does not contain the spinor representation of 
$SO(10)$, but it contains the spectrum of the standard model. 
Indeed, as pioneered in Ref.~\cite{Witten:1984dg}, several standard-like models from the $SO(32)$ heterotic string 
theory are constructed on toroidal orbifold~\cite{Giedt:2003an,Choi:2004wn,Nilles:2006np,RamosSanchez:2008tn}, torus with magnetic fluxes~\cite{Abe:2015mua,Abe:2016eyh}, elliptically fibered CY manifolds with stable vector bundles~\cite{Blumenhagen:2005zg} given by the spectral cover 
construction~\cite{Friedman:1997ih,Friedman:1997yq} and smooth CY threefolds with line bundles~\cite{Nibbelink:2015vha}. 
However, the $SO(32)$ heterotic line bundle models on smooth CY threefolds have not been fully explored.

S- and T-dualities tell us that the $SO(32)$ heterotic line bundle models correspond to intersecting D6-brane models in type 
IIA string theory, where several stacks of branes directly lead to the minimal supersymmetric 
standard model (MSSM) and/or Pati-Salam model~\cite{Ibanez:2001nd,Cvetic:2001tj,Blumenhagen:2005mu}.\footnote{For 
more details, see e.g. Ref.~\cite{Honecker:2016gyz} and references therein.} 
It therefore motivates us to search for standard-like models from the $SO(32)$ heterotic string theory on 
CY threefolds without an intermediate GUT.\footnote{For the model building realizing the $SU(5)$-like spectrum, we refer to Ref.~\cite{Nibbelink:2015vha}.} 
To obtain realistic three-generation models, we consider multiple line bundles in the Cartan directions of $SO(32)$ 
rather than non-abelian bundles. 
These line bundles allow us to decompose the $SO(32)$ gauge group and compute the net chiral asymmetries 
of quarks and leptons, taking into account that the choice of line bundles is constrained by the consistency conditions such as 
anomaly cancellation conditions, masslessness conditions of the hypercharge gauge boson, 
supersymmetric conditions, and so-called K-theory condition. 
It turns out that these theoretical and phenomenological requirements indicate the enhancement of the standard model gauge group 
to the Pati-Salam one.  
To break the Pati-Salam gauge symmetry, 
we concentrate on quotient complete intersection Calabi-Yau manifolds (CICYs)~\cite{Candelas:1987kf,Candelas:1987du} 
by a similar argument as in $E_8\times E_8$ heterotic line bundle models~\cite{Anderson:2011ns,Anderson:2012yf}, 
where one can introduce Wilson lines into the internal components of $U(1)$s because of the existence of a freely-acting discrete symmetry group. 
We systematically search for standard-like models on several CICYs with multiple line bundles, 
where the number of K\"ahler moduli is restricted to $1\leq h^{1,1}\leq 5$, 
with $h^{1,1}$ being the hodge number of CICYs. 
We find that such restrictive line bundles and certain Wilson lines lead to the standard-like models where the gauge group consists of 
$SU(3)_C\times SU(2)_L \times U(1)_Y \times U(1)_{B-L}$ via the Pati-Salam-like symmetry. 
The spectrum in the visible sector contains the three generations of quarks and leptons without chiral exotics, and Higgs doublets 
appear as the vector-like particles with respect to the standard model gauge group with an extra $U(1)_{B-L}$. 
The gauge symmetries of the low-energy effective action allow for the 
perturbative Yukawa couplings of quarks and leptons.

The remainder of this paper is organized as follows. 
We first briefly review several consistency conditions on the basis of the low-energy 
effective action of the heterotic string theory in Sec.~\ref{subsec:2_1}. 
In Sec.~\ref{subsec:2_2}, we present the group decomposition of $SO(32)$ gauge symmetry employing multiple line bundles 
along the line of Ref.~\cite{Abe:2015mua}. 
Even if line bundles satisfy the consistency conditions, they are further constrained by 
the masslessness conditions of the $U(1)_Y$ gauge boson, originating from 
the couplings between the $U(1)_Y$ gauge boson and closed string axions as discussed in Sec.~\ref{subsec:2_3}. 
A search for concrete three-generation standard-like models on several quotient CICYs 
indicates in Sec.~\ref{sec:3} that the visible gauge group consists of $SU(3)_C\times SU(2)_L\times U(1)_Y\times U(1)_{B-L}$ 
as a result of multiple line bundles and certain Wilson lines. 
Other $U(1)$ gauge bosons become massive through the Green-Schwarz mechanism. 
The obtained spectrum contains the three generations of quarks and leptons, 
vector-like Higgs and extra vector-like particles with respect to $SU(3)_C\times SU(2)_L\times U(1)_Y\times U(1)_{B-L}$. 
%The obtained chiral spectrum except for the singlets is just the standard model one. 
Extra $U(1)$ gauge symmetries including $U(1)_{B-L}$ control the Yukawa couplings among the elementary particles 
and higher-dimensional operators causing the proton decay. 
Finally, Sec.~\ref{sec:con} is devoted to the conclusion.

%%%%%%%%%%%%%%%%%%%%%%%%%%%%%%%%%%%%%%%%%%%%%%%%%%%%%%%%%%%%%%%%%%%%%%%%%%%%%%%%%%%%%%%%%%%%%%%%%%%%%%%%
\section{Setup}
\label{sec:2}
\subsection{Consistency conditions in the low-energy effective action}
\label{subsec:2_1}
We briefly review the low-energy effective action of the $SO(32)$ heterotic string theory on CY manifolds with 
multiple line bundles. (For more details, we refer to Refs.~\cite{Blumenhagen:2005ga,Blumenhagen:2005pm,Polchinsky}.)  
At the order of $\alpha^\prime$, 
the bosonic part of the low-energy effective action is given by 
\begin{align}
S_{\rm bos}&=\frac{1}{2\kappa_{10}^2}\int_{M^{(10)}} 
e^{-2\phi_{10}} \left[ R+4d\phi_{10} \wedge  \ast 
d\phi_{10} -\frac{1}{2}H\wedge \ast H \right] 
\nonumber\\
&-\frac{1}{2g_{10}^2} \int_{M^{(10)}} e^{-2\phi_{10}} 
{\rm tr}(F\wedge \ast F)
-\sum_s N_s T_5 \int_{M^{(10)}} B^{(6)} \wedge \delta(\gamma_s), 
\label{eq:heterob}
\end{align}
where $\phi_{10}$ denotes the dilaton. 
Trace of $F$ and $R$, ``tr'', is taken in the fundamental representations 
of $SO(32)$ and $SO(1,9)$, $\kappa_{10}$ and $g_{10}$ are 
the gravitational and gauge couplings normalized as $2\kappa_{10}^2=(2\pi)^7(\alpha')^4$ 
and $g_{10}^2=2(2\pi)^7(\alpha')^3$. 
Here, we include the Wess-Zumino term in the presence of heterotic five-branes with tension $T_5=((2\pi)^5(\alpha')^3)^{-1}$. 
Heterotic (anti) five-branes wrapping the holomorphic two-cycles $\gamma_s$ 
correspond to the positive (negative) $N_s$ and the Poincar\'{e} dual four-form of $\gamma_s$ is 
represented as $\delta (\gamma_s)$. 
$H=dB^{(2)} -\frac{\alpha^{'}}{4}(w_{\rm YM} -w_{L})$ involves the gauge and gravitational Chern-Simons three-forms, $w_{\rm YM}$ and $w_{L}$. 
Note that Kalb-Ramond two-form $B^{(2)}$ is related to $B^{(6)}$ under the ten-dimensional 
Hodge duality, namely $\ast dB^{(2)}=e^{2\phi_{10}}dB^{(6)}$. 

Throughout this paper, we consider the following internal gauge bundle
\begin{align}
W=\bigoplus_{a=1}^M L_a,
\label{eq:WLa}
\end{align}
where $L_a$ are the multiple line bundles, each with structure group $U(1)$. 
The concrete embedding of $U(1)$ into $SO(32)$ is discussed later. 
The inclusion of such line bundles breaks $SO(32)$ into the four-dimensional gauge group $G$ 
and the adjoint representation of $SO(32)$ is decomposed as
\begin{align}
496\rightarrow \bigoplus_p (R_p, C_p),
\end{align}
where $R_p$ and $C_p$ stand for certain representations of $G$ and $W$. 
Given these line bundles, we can construct standard-like models only if 
internal gauge bundles satisfy several consistency conditions 
which are enumerated as follows.

First of all, when we denote the internal background field strengths as $\bar{F}$ and $\bar{R}$, 
the Bianchi identity of the Kalb-Ramond field $B^{(6)}$ 
\begin{align}
d(e^{2\phi_{10}}\ast dB^{(6)}) =-\frac{\alpha^{'}}{4} 
\left({\rm tr}\bar{F}^2 -{\rm tr}{\bar R}^2 -
4(2\pi)^2 \sum_s N_s \delta(\gamma_s) \right)
\end{align}
constrains the abelian bundles as
\begin{align}
{\rm ch}_2(W)+c_2(T{\cal M})=
\sum_s N_s \delta(\gamma_s), 
\label{eq:tad}
\end{align}
where ${\rm ch}_2(W)$ and $c_2(T{\cal M})$ are the second Chern character and second Chern class of 
$W$ and the tangent bundle of the CY manifold ${\cal M}$ respectively. 
To keep the stability of the system, we require the vanishing anti heterotic five-branes, namely 
\begin{align}
{\rm ch}_2(W)+c_2(T{\cal M})\geq [0],
\label{eq:tad2}
\end{align}
in cohomology.

Next, since the spinorial representation appears in the first excited mode in the heterotic string~\cite{Gross:1985fr,Gross:1985rr}, 
we require that the first Chern class of the total gauge bundle lies in second even integral cohomology basis of the CY manifold:
\begin{align}
c_1(W)=\sum_a n_a c_1(L_a) \in H^2 ({\cal M}, 2\mathbb{Z}),
\end{align}
where $n_a$ are the integers depending on the embedding of line bundles into $SO(32)$. 
Such a condition is also related to the K-theory condition~\cite{Witten:1998cd,Uranga:2000xp} in the S-dual type I superstring theory.

Finally, the nonvanishing $U(1)$ field strengths on the CY manifold are constrained by 
the supersymmetric condition. From the supersymmetric transformations of the gauginos, 
the internal field strengths have to obey 
\begin{align}
\bar{F}_{ij}=\bar{F}_{\bar{i}\bar{j}}=0,\qquad
g^{i\bar{j}}\bar{F}_{i\bar{j}}=0,
\end{align}
implying that the vector bundles should be holomorphic. 
It is known that the constraint equation for the $(1,1)$ form of $F$ is called the Hermitian Yang-Mills 
equation. It can be solved when holomorphic bundles satisfy the following condition
\begin{align}
\mu(L_a)=\frac{1}{2l_s^4}\int_{\cal M} J\wedge J\wedge c_1(L_a)
+\frac{{\cal V}}{6\pi s}\int_{\cal M} \biggl[ c_1^3(L_a)+\frac{1}{4}c_1(L_a)\wedge c_2(T{\cal M}) \biggl]=0,
\label{eq:Dterm}
\end{align}
for all $a$. $J=l_s^2\sum_{i=1}^{h^{1,1}}t^iw_i$ is the K\"ahler form expressed in terms of a basis $w_i$ of $H^{1,1}({\cal M}, \mathbb{R})$, 
where $t^i$ denote the K\"ahler moduli normalized by the string length $l_s=2\pi \sqrt{\alpha^\prime}$. 
Now, we include the string loop corrections~\cite{Blumenhagen:2005ga,Blumenhagen:2005pm} 
characterized by the volume of CY ${\cal V}=\frac{1}{6}d_{ijk}t^it^jt^k$ and the dilaton $s={\cal V}/(2\pi g_s^2)$. 
$d_{ijk}$ stand for the triple intersection numbers of CY manifolds, namely $d_{ijk}=\int_{\cal M}w_i \wedge w_j\wedge w_k$. 
Hence, the first Chern class of each line bundle should be properly chosen such that the 
moduli spaces of K\"ahler moduli reside in the supergravity reliable domain $t^i> 1$ 
in string units. 
Otherwise, the nonvanishing $D$-terms appear in the four-dimensional supergravity action. 
%Note that the above constraint is only valid at the tree-level as discussed in Ref.~\cite{Blumenhagen:2005ga,Blumenhagen:2005pm}. 
In this paper, we restrict ourselves to line bundles satisfying all the consistency conditions presented so far.

In this approach, we can calculate the net-number of chiral massless fermions. 
When we denote the internal bundle ${\cal C}_p$ to each $C_p$, the left-handed and the right-handed fermionic zero-modes 
in $R_p$ are counted by the Dolbeault cohomology $H^1 ({\cal M}, {\cal C}_p)$ and $H^2 ({\cal M}, {\cal C}_p)$ respectively. 
Since the $\mu$-stable bundles ($\mu(L_a)=0$) give the vanishing zeroth and third cohomology $H^0({\cal M}, {\cal C}_p)=H^3({\cal M}, {\cal C}_p)=0$\footnote{As commented in Ref.~\cite{Anderson:2012yf}, non-trivial line bundles $L_a$ have ${\rm dim}(H^0({\cal M}, L_a))=0$ if $\mu(L_a)<0$ 
and also ${\rm dim}(H^3({\cal M}, L_a))=0$ if $\mu (L_a)>0$. Such positive and negative slopes $\mu (L_a)$ exist in the 
K\"ahler moduli space, since we focus on the case with $\mu(L_a)=0$. 
Hence, we obtain ${\rm dim}(H^0({\cal M}, L_a))={\rm dim}(H^3({\cal M}, L_a))=0$ according to the theorem in Ref.~\cite{Kobayashi}.} 
and $H^2({\cal M}, {\cal C}_p)\simeq H^1({\cal M}, {\cal C}_p^\ast)$ by Serre duality, 
Hirzebruch-Riemann-Roch theorem tells us that the net-number of chiral massless fermions 
(chiral supermultiplets) is counted by the 
corresponding Euler number
\begin{align}
\chi ({\cal M}, {\cal C}_p)&=\sum_{i=0}^3 (-1)^i{\rm dim}(H^i({\cal M}, {\cal C}_p))=
-{\rm dim}(H^1({\cal M}, {\cal C}_p))+{\rm dim}(H^1({\cal M}, {\cal C}_p^\ast))
\nonumber\\
&=
\int_{\cal M}\biggl[{\rm ch}_3({\cal C}_p)+\frac{1}{12}c_2(T{\cal M})c_1({\cal C}_p)\biggl].
\label{eq:chi}
\end{align}
Note that we now consider the complex representation $C_p$. 
When some bundles in Eq.~(\ref{eq:WLa}) are trivial bundles ${\cal O}_{\cal M}$, 
the $U(1)$ gauge symmetries are enhanced to non-abelian ones as shown in Sec.~\ref{sec:3} 
and corresponding cohomology becomes 
${\rm dim}(H^0({\cal M}, {\cal O}_{\cal M}))={\rm dim}(H^3({\cal M}, {\cal O}_{\cal M}))=1$ and ${\rm dim}(H^1({\cal M}, {\cal O}_{\cal M}))={\rm dim}(H^2({\cal M}, {\cal O}_{\cal M}))=0$. 
This is because the zero-modes of the Dirac operator are the $(0,0)$ and $(0,3)$ forms under the Dolbeault operator $\bar{\partial}$ 
on manifolds of SU(3) holonomy. For more details we refer to Ref.~\cite{Green:1987mn}.

In this way, proper internal line bundles have the potential to yield three generations of chiral fermions for 
a large class of CY manifolds, although, in the standard embedding scenario, 
three-generation models are restricted to specific CY manifolds with small hodge numbers.

%%%%%%%%%%%%%%%%%%%%%%%%%%%%%%%%%%%%%%%%%%%%%%%%%%%%%%%%%%%%%%%%%%%%%%%%%%%%%%%%%%%%%%%%%%%%%%%%%%%%%%
\subsection{Matter content and group decomposition}
\label{subsec:2_2}
In the following, we proceed to extract standard-like models from the $SO(32)$ 
heterotic string theory. 
Along the line of Ref.~\cite{Abe:2015mua}, we first decompose the $SO(32)$ gauge group as
\begin{align}
SO(32) &\rightarrow SO(12) \times SO(20), 
\nonumber\\
496 &\rightarrow (1, 190) + (12_v, 20_v)
+ (66, 1),
\end{align}
where $SO(12)$ and $SO(20)$ are further decomposed by the insertion of line bundles,
\begin{align}
SO(12) 
&\rightarrow SO(8) \times SU(2)_L \times U(1)_1 
\rightarrow SU(4)_C \times U(1)_2 \times SU(2)_L \times U(1)_1 
\nonumber\\
&\rightarrow SU(3)_C \times U(1)_3 \times U(1)_2 
\times SU(2)_L \times U(1)_1,
\nonumber\\
SO(20) &\rightarrow U(1)_4 \times \cdots \times U(1)_{13}.
\end{align}
Under the above decomposition, we take the Cartan directions of 
$SO(32)$, $H_a (a=1,2,\cdots, 16)$, as $H_1-H_2$ and $H_1+H_2-2H_3$ for $SU(3)_C$ 
and $H_5-H_6$ for $SU(2)_L$. 
Other $U(1)$ directions are chosen as
\begin{align}
U(1)_1&: (0,0,0,0,1,1;0,\cdots,0),
\nonumber\\
U(1)_2&: (1,1,1,1,0,0;0,\cdots,0),
\nonumber\\
U(1)_3&: (1,1,1,-3,0,0;0,\cdots,0),
\nonumber\\
U(1)_4&: (0,0,0,0,0,0;1,0,\cdots,0),
\nonumber\\
U(1)_5&: (0,0,0,0,0,0;0,1,0,\cdots,0),
\nonumber\\
&\vdots
\nonumber\\
U(1)_{13}&: (0,0,0,0,0,0;0,\cdots,0,1),
\label{eq:U(1)cartan}
\end{align}
in the basis $H_a$ and $SO(32)$ roots are 
$(\underline{\pm 1, \pm 1, 0,\cdots, 0})$ under $H_a$, $(a=1,\cdots, 16)$. 
The underline represents all the possible permutations. 

\begin{table}[htb]
\centering
  \begin{tabular}{|c|c|c|c|c|} \hline
    $SO(12)$ & $\left(SU(4)_C\times SU(2)_L\right)_{U(1)_1,U(1)_2}$ &  $\left(SU(3)_C\times SU(2)_L\right)_{U(1)_1,U(1)_2,U(1)_3}$ 
    & $U(1)_Y$ & Matter
\\ \hline \hline
    $66$ & $(15,1)_{0,0}$ & $(3,1)_{0,0,4}$ & $2/3$ & $\bar{u}_{R_1}^c$ 
      \\ 
           &   & $(\bar{3},1)_{0,0,-4}$ & $-2/3$ & $u_{R_1}^c$ 
      \\ \hline
           & $(6,1)_{0,2}$ & $(\bar{3},1)_{0,2,2}$  & $1/3$ & $d_{R_1}^c$ 
      \\ 
           &   & $(3,1)_{0,2,-2}$ & $-1/3$ & $\bar{d}_{R_2}^c$ 
      \\ \hline
           & $(6,1)_{0,-2}$ & $(\bar{3},1)_{0,-2,2}$ & $1/3$ & $d_{R_2}^c$ 
      \\ 
           &   &  $(3,1)_{0,-2,-2}$ & $-1/3$ & $\bar{d}_{R_1}^c$
      \\ \hline 
           & $(4,2)_{1,1}$ & $(3,2)_{1,1,1}$ & $1/6$ & $Q_1$ 
      \\ 
           &   & $(1,2)_{1,1,-3}$ & $-1/2$ & $L_1$ 
      \\ \hline
           & $(\bar{4},2)_{1,-1}$ & $(\bar{3},2)_{1,-1,-1}$ & $-1/6$ & $\bar{Q}_2$ 
      \\ 
           &   & $(1,2)_{1,-1,3}$ & $1/2$ & $\bar{L}_2$ 
      \\ \hline
           & $(4,2)_{-1,1}$ & $(3,2)_{-1,1,1}$ & $1/6$ & $Q_2$ 
      \\ 
           &   & $(1,2)_{-1,1,-3}$ & $-1/2$ & $L_2$ 
      \\ \hline
           & $(\bar{4},2)_{-1,-1}$ & $(\bar{3},2)_{-1,-1,-1}$ & $-1/6$ & $\bar{Q}_1$ 
      \\ 
           &   & $(1,2)_{-1,-1,3}$ & $1/2$ & $\bar{L}_1$ 
      \\ \hline
           & $(1,1)_{2,0}$  & $(1,1)_{2,0,0}$ & $0$ & $n_1$ 
      \\ \hline
           & $(1,1)_{-2,0}$  & $(1,1)_{-2,0,0}$ & $0$ & $\bar{n}_1$ 
      \\ \hline                                              
  \end{tabular}
  \caption{Matter content from the adjoint representation of $SO(12)$, where the hypercharge $U(1)_Y$ is identified as $U(1)_3/6$. 
  Here, the subscript indices label the $U(1)$ charges.}
  \label{tab:2_2}
\end{table}

Under this Cartan basis, the adjoint representation of $SO(12)$, $66$, 
involves all the candidates of the standard model particles, except for the right-handed leptons as 
summarized in Table~\ref{tab:2_2}. 
Hence, it motivates us to further decompose $SO(20)$ into multiple $U(1)$s. 
When $SO(20)$ is decomposed into all the $U(1)$s, 
the candidates of right-handed leptons appear from 
the vector representation and the singlet of $SO(12)$, $12_v$ and 1, 
\begin{align}
(12_v, 20_v) &\rightarrow 
\left\{ 
\begin{array}{l}
L_3^a =(1,2)_{1,0,0;-1_{(a)}}\\
L_4^a =(1,2)_{-1,0,0;-1_{(a)}}\\
u_{R_2}^{c\,\,a}=({\bar 3},1)_{0,-1,-1;-1_{(a)}}\\
d_{R_3}^{c\,\,a}=({\bar 3},1)_{0,-1,-1;1_{(a)}}\\
e_{R_1}^{c\,\,a}=(1,1)_{0,-1,3;1_{(a)}}\\
n_{2}^{c\,\,a}=(1,1)_{0,-1,3;-1_{(a)}}\\
\end{array}
\right.
,\,(a=4,5,\cdots, 13),
\nonumber\\
(1, 190) &\rightarrow 
\left\{ 
\begin{array}{l}
e_{R_2}^{c\,\,ab}=(1,1)_{0,0,0;1_{(a)},1_{(b)}}\\
n_{3}^{c\,\,ab}=(1,1)_{0,0,0;1_{(a)},-1_{(b)}}\\
%n_{3}^{c\,\,4a}=(1,1)_{0,0,0;1,\underline{-1,0,\cdots,0}}\\
%n_{3}^{c\,\,ab}=(1,1)_{0,0,0;0,1\underline{1,-1,0,\cdots,0}}\\
\end{array}
\right.
,\,(a,b=4,5,\cdots, 13,\,a< b),
\end{align}
where the subscript indices label the $U(1)_{1,2,3}$ and non-zero $U(1)_{a,b}$ charges.
This decomposition results in the candidates of right-handed quarks, charged-leptons and/or Higgs.

To realize the correct hypercharge, we redefine the hypercharge as 
\begin{align}
U(1)_Y=\frac{1}{6} \left( U(1)_3 +3\sum_{c=4}^{13}U(1)_c\right),
\label{eq:U(1)Y}
\end{align}
and the matter content and associated cohomology are then summarized in Table~\ref{tab:2_2_2}. 
It is notable that we just decompose $SO(20)$ into tenth $U(1)$s by inserting line bundles into all the Cartan directions. 
If the first Chern numbers of certain $U(1)$s are vanishing or correlated with each other, these $U(1)$ 
gauge symmetries are enhanced to non-abelian ones as demonstrated in Sec.~\ref{sec:2_4}.

\begin{table}[htb]
\centering
  \begin{tabular}{|c|c|c|} \hline
    Matter & Repr. & Cohomology  \\ \hline \hline
    $Q_1$ & $(3,2)_{1,1,1}$ & $H^\ast({\cal M},  L_1\otimes L_2 \otimes L_3)$ 
     \\
    $Q_2$ & $(3,2)_{-1,1,1}$ & $H^\ast({\cal M}, L_1^{-1}\otimes L_2 \otimes L_3)$  \\ \hline
    $L_1$ & $(1,2)_{1,1,-3}$ & $H^\ast({\cal M},  L_1\otimes L_2 \otimes L_3^{-3})$ \\
    $L_2$ & $(1,2)_{-1,1,-3}$ & $H^\ast({\cal M}, L_1^{-1}\otimes L_2 \otimes L_3^{-3})$  \\ \hline
    $L_3^a$ & $(1,2)_{1,0,0;-1_{(a)}}$ & $H^\ast({\cal M},  L_1\otimes L_a^{-1})$ \\
    $L_4^a$ & $(1,2)_{-1,0,0;-1_{(a)}}$ & $H^\ast({\cal M}, L_1^{-1}\otimes L_a^{-1})$  \\ \hline 
    $u_{R_1}^c$ & $({\bar 3},1)_{0,0,-4}$ & $H^\ast({\cal M},  L_3^{-4})$ \\
    $u_{R_2}^{c\,\,a}$ & $({\bar 3},1)_{0,-1,-1;-1_{(a)}}$ & $H^\ast({\cal M}, L_2^{-1} \otimes L_3^{-1}\otimes L_a^{-1})$  \\ \hline
    $d_{R_1}^c$ & $({\bar 3},1)_{0,2,2}$ & $H^\ast({\cal M},  L_2^2 \otimes L_3^2)$ \\
    $d_{R_2}^c$ & $({\bar 3},1)_{0,-2,2}$ & $H^\ast({\cal M},  L_2^{-2} \otimes L_3^2)$ \\
    $d_{R_3}^{c\,\,a}$ & $({\bar 3},1)_{0,-1,-1;1_{(a)}}$ & $H^\ast({\cal M}, L_2^{-1} \otimes L_3^{-1}\otimes L_a)$  \\ \hline
    $e_{R_1}^{c\,\,a}$ & $(1,1)_{0,-1,3;1_{(a)}}$ & $H^\ast({\cal M}, L_2^{-1} \otimes L_3^3\otimes L_a)$  \\ 
    $e_{R_2}^{c\,\,ab}$ & $(1,1)_{0,0,0;1_{(a)},1_{(b)}}$ & $H^\ast({\cal M}, L_a\otimes L_b)$  \\ \hline    
    $n_1$ & $(1,1)_{2,0,0}$ & $H^\ast({\cal M},  L_1^{2})$ \\    
    $n_{2}^{c\,a}$ & $(1,1)_{0,-1,3;-1_{(a)}}$ & $H^\ast({\cal M}, L_2^{-1} \otimes L_3^3\otimes L_a^{-1})$  \\ 
    $n_{3}^{c\,ab}$ & $(1,1)_{0,0,0;1_{(a)},-1_{(b)}}$ & $H^\ast({\cal M}, L_a\otimes L_b^{-1})$ \\ \hline  
  \end{tabular}
  \caption{Massless spectrum and corresponding cohomology. In the second column, 
  the subscript indices label the $U(1)_{1,2,3}$ and non-zero $U(1)_{a,b}$ charges.}
  \label{tab:2_2_2}
\end{table}

%%%%%%%%%%%%%%%%%%%%%%%%%%%%%%%%%%%%%%%%%%%%%%%%%%%%%%%%%%%%%%%%%%%%%%%%%%%%%%%%%%%%%%%%%%%%%%%%%%%
\subsection{$U(1)_Y$ masslessness conditions}
\label{subsec:2_3}
Before searching for three generations of quarks and leptons, 
let us discuss the couplings between string axions and the $U(1)_Y$ gauge boson. 
Such axionic couplings will cause the mass term of the $U(1)_Y$ gauge boson, 
even when gauge bundles satisfy the consistency conditions presented in Sec.~\ref{subsec:2_1}. 
The authors of Refs.~\cite{Blumenhagen:2005ga,Blumenhagen:2005pm} pointed out that 
background gauge fluxes induce the couplings between 
string axions and $U(1)$ gauge bosons through the Green-Schwarz term~\cite{Green:1984bx,Ibanez:1986xy},
\begin{align}
S_{\rm GS}=\frac{1}{24(2\pi)^5\alpha'}\int B^{(2)}\wedge X_8,
\label{eq:GS}
\end{align}
where
\begin{align}
X_8=\frac{1}{24}{\rm Tr}F^4-\frac{1}{7200}({\rm Tr}F^2)^2-
\frac{1}{240}({\rm Tr}F^2)({\rm tr}R^2) +\frac{1}{8}{\rm tr}R^4 
+\frac{1}{32}({\rm tr}R^2)^2.
\label{eq:X8}
\end{align}
Now ``Tr'' is taken in the adjoint representation of $SO(32)$. 

Under the expansion of the Kalb-Ramond field in a basis of $H^2({\cal M}, \mathbb{Z})$,
\begin{align}
&B^{(2)} =b_0^{(2)} +l_s^2\sum_{i=1}^{h^{1,1}} b_i^{(0)} w_i, 
\nonumber\\
&B^{(6)} =l_s^6b_0^{(0)}{\rm vol}_6 
+l_s^4\sum_{i=1}^{h^{1,1}} b_i^{(2)} \hat{w}^i, 
\end{align}
where $\hat{w}^i$ are the Hodge dual four-forms of the 
K\"ahler forms $w_i$ and ${\rm vol}_6$ denotes the volume form of the CY manifold, 
$b_i^{(0)}$ and $b_0^{(0)}$ represent the model-dependent and -independent axions respectively. 
Those are connected with two-forms, $b_i^{(2)}$ and $b_0^{(2)}$ under the ten-dimensional 
hodge duality $\ast dB^{(2)}=e^{2\phi_{10}}dB^{(6)}$.

On the line bundle background, the $U(1)_a$ gauge field strengths are decomposed as
\begin{align}
F_a=f_a+\bar{f}_a,
\end{align}
where $f_a$ are four-dimensional parts and
\begin{align}
\bar{f}_a=2\pi \sum_{i=1}^{h^{1,1}} m_a^{i} w_i
\end{align}
denote the background gauge fluxes in a basis of $H^{2}({\cal M}, \mathbb{Z})$. 
Here, $m_a^{i}$ are the integers subject to the Dirac quantization condition and 
we assume that the first Chern numbers of all the $U(1)$s 
are independent, otherwise these $U(1)$ gauge symmetries are enhanced to non-abelian ones. 
Such a gauge background gives rise to the axionic couplings 
through the Green-Schwarz term,
\begin{align}
\frac{1}{3(2\pi )^3 l_s^2} 
\int b_0^{(2)} \wedge &\Bigl[ 
{\rm tr}T_1^4{\bar f}_1^3f_1 +
\left({\rm tr}T_2^4{\bar f}_2^3 +
3({\rm tr}T_2^2T_3^2){\bar f}_2{\bar f}_3^2 +
({\rm tr}T_2T_3^3){\bar f}_3^3 
\right) f_2 
\nonumber\\
&+\left(
{\rm tr}T_3^4{\bar f}_3^3 +
3({\rm tr}T_2T_3^3){\bar f}_2{\bar f}_3^2 +
3({\rm tr}T_2^2T_3^2){\bar f}_2^2{\bar f}_3
\right) f_3 
+\sum_{c=4}^{13}{\rm tr}T_c^4{\bar f}_c^3f_c\Bigl],
\end{align}
for the model-independent axion\footnote{Note that the curvature terms in Eq.~(\ref{eq:X8}) are irrelevant to the masslessness 
conditions of $U(1)_Y$.} and 
\begin{align}
&\frac{1}{l_s^2} 
\int b_{i}^{(2)} \wedge \left(\sum_{a=1}^{13}{\rm tr}(T_a^2)m_{a}^{i}\right) f_a,
\end{align}
for the model-dependent axions.\footnote{For the detailed derivation, see Ref.~\cite{Abe:2015mua}.} 
Here, $T_a$ denote the $U(1)_a$ generators whose directions are chosen as in Eq.~(\ref{eq:U(1)cartan}). 
These Stueckelberg couplings yield the mass terms of the $U(1)_a$ gauge bosons as mentioned before.

Recalling the fact that the hypercharge $U(1)_Y$ is defined 
as $U(1)_Y=\frac{1}{6}(U(1)_3 +3\sum_{c=4}^{13}U(1)_c)$, 
the masslessness of the $U(1)_Y$ gauge field is ensured when
\begin{align}
&{\rm tr}(T_3^4) d_{ijk}m_3^{i}m_3^{j}m_3^{k} 
+3{\rm tr}(T_2T_3^3) d_{ijk}m_2^{i}m_3^{j}m_3^{k}
+3{\rm tr}(T_2^2T_3^2) d_{ijk}m_2^{i}m_2^{j}m_3^{k}
\nonumber\\
&+3\sum_{c=4}^{13} {\rm tr}(T_c^4)d_{ijk}m_c^{i}m_c^{j}m_c^{k}=0
\label{eq:massless1}
\end{align} 
and
\begin{align}
&{\rm tr}(T_3^2)m_3^{i} +3\sum_{c=4}^{13} {\rm tr}(T_c^2)m_c^{i}=0
\label{eq:massless2}
\end{align} 
are satisfied simultaneously.

%%%%%%%%%%%%%%%%%%%%%%%%%%%%%%%%%%%%%%%%%%%%%%%%%%%%%%%%%%%%%%%%%%%%%%%%%%%%%%%%%%%%%%%%%%%%%%%%%%%%%%%%%%
\subsection{Pati-Salam-like models}
\label{sec:2_4}
The simplest way to satisfy all the requirements is to set
\begin{align}
m_5^i=-m_4^i,\qquad
m_3^i=m_d^i=0,
\label{eq:ansatz}
\end{align}
for $d=6,\cdots,13$ and $i=1,2,\cdots, h^{1,1}$ and non-zero values for other line bundles. 
Under the above ansatz, one can easily satisfy the K-theory condition
\begin{align}
c_1(W)=2c_1(L_1)+4c_1(L_2)+\sum_{c=4}^{13}c_1(L_c)=0~({\rm mod}\,2), 
\end{align}
and the $U(1)_Y$ masslessness conditions in Eqs.~(\ref{eq:massless1}) and (\ref{eq:massless2}). 
Here and in what follows, the traces of $U(1)$ generators are taken as ${\rm tr}(T_{1})=2$, ${\rm tr}(T_{2})=4$, ${\rm tr}(T_{c})=1$, 
where $U(1)_1$ and $U(1)_2$ are embedded into $U(2)_L$ and $U(4)_C$ gauge groups respectively. 
Note that the gauge symmetries $SU(2)_L$ and $SU(4)_C$ of the Pati-Salam model are embedded as 
$SU(2)_L \subset U(2)_L$ and $SU(4)_C \subset U(4)_C$. 
These first Chern classes are of the form
\begin{align}
c_1(L_a)=\sum_{i=1}^{h^{1,1}}m_a^i w_i,
\end{align}
where the quantized fluxes $m_a^i$ are the integers as mentioned before. 
However, the above choice of line bundles results in the Pati-Salam-like gauge group plus hidden gauge group,
\begin{align}
SO(32)\rightarrow SU(4)_C\times SU(2)_L \times SU(2)_R \times \Pi_{c=1}^2U(1)_c \times U(1)^\prime \times SO(16),
\end{align}
where $SU(3)_C$ is enhanced to $SU(4)_C$ due to the vanishing $c_1(L_3)$. 
The correlated first Chern classes $m_5^i=-m_4^i$ also lead to the 
gauge enhancement from $U(1)_{4,5}$ to $SU(2)_R\otimes U(1)^\prime$ 
whose Cartan directions are taken as $H_4+H_5$ for $SU(2)_R$ and $H_4-H_5$ for $U(1)^\prime \simeq S(U(1)_4\times U(1)_5)$. 
For our purpose, we search for the three-generation models using $S(U(1)_4\times U(1)_5)$ rather than $U(1)^\prime$ 
in the latter analysis, where two $U(1)_{4,5}$ charge vectors 
${\bm q}=(q_4, q_5)$ and ${\bm q}^\prime=(q_4^\prime, q_5^\prime)$ are identified if ${\bm q}-{\bm q}^\prime \in \mathbb{Z}\times (1,1)$ because of $L_5\simeq L_4^{-1}$. 
Other $U(1)$s, especially $SO(2)$s, are enhanced to hidden $SO(16)$ due to the vanishing 
first Chern classes. 
As a result, the remaining gauge symmetry on ${\cal M}$ is similar to the Pati-Salam model. 
%The net-number of chiral supermultiplet in Table~\ref{tab:1} is counted through Eq.~(\ref{eq:chi}). 
It is remarkable that the above discussion presented so far is irrelevant to the underlying CY geometries thanks to 
the constraints on the line bundles~(\ref{eq:ansatz}). 
Although it is possible to directly derive the standard model gauge group with hypercharge flux breaking, 
we require a concrete model-by-model search. We therefore leave the detailed analysis for a future work.

When the vacuum expectation values of singlet fields under the Pati-Salam-like gauge group are taken zero, 
the number of massive $U(1)$s is given by the rank of the following mass matrix in units of 
string length $l_s=2\pi\sqrt{\alpha'}$~\cite{Blumenhagen:2005ga,Blumenhagen:2005pm},
\begin{align}
M_{ai}=\left\{
\begin{array}{c}
2\pi\,{\rm tr}(T_a)m_a^i\qquad {\rm for}\,i=1,2,\cdots, h^{1,1}
\\
d_{jkl}\,{\rm tr}(T_a)\biggl[\frac{1}{6}m_a^jm_a^km_a^l+\frac{1}{24}m_a^jc_2^{kl}(T{\cal M})\biggl]
\qquad {\rm for}\,i=0
\end{array}
\right..
\label{eq:Mai}
\end{align}
The first line is derived from the Stueckelberg couplings between 
the model-dependent axions and $U(1)_a$ gauge bosons, whereas the second line is originating 
from the model-independent axion associated with the dilaton field. 
It turns out that when $h^{1,1}=1$, the maximum rank of mass matrix (\ref{eq:Mai}) is $2$ in which 
at least one of the three U(1)s remains massless due to the correlated Chern numbers~(\ref{eq:ansatz}). 
When $h^{1,1}\geq 2$, it is possible to 
consider three massive $U(1)$ gauge bosons at the compactification scale. 
Note that the Cartan direction of $U(1)_Y=\frac{1}{6}\left( U(1)_3+3U(1)_4+3U(1)_5\right)$ remains massless and 
the remaining $U(1)_3$ could be identified with $U(1)_{B-L}$ after the Pati-Salam symmetry breaking. 
%The matter contents on ${\cal M}$ is summarized in Table~\ref{}. 

In addition to the constraints on the line bundles~(\ref{eq:ansatz}), 
we restrict ourselves to line bundles satisfying the $D$-term conditions~(\ref{eq:Dterm}) and 
the stability condition~(\ref{eq:tad2}) simplified as 
\begin{align}
2{\rm ch}_2(L_1)+4{\rm ch}_2(L_2)+{\rm ch}_2(L_4)+{\rm ch}_2(L_5)+c_2(T{\cal M})\geq [0],
\label{eq:anomalysec3}
\end{align}
in cohomology, where ${\rm ch}_2(L_a)=\frac{1}{2}c_1^2(L_a)=\frac{1}{2}d_{ijk}m_a^jm_a^k \tilde{w}^i$ 
with $c_1^i(L_a)=m_a^i$. 
Note that we use the relations ${\rm tr}(T_1^2)=2$, ${\rm tr}(T_2^2)=4$ and ${\rm tr}(T_{4,5}^2)=1$. 
In the light of $D$-term conditions~(\ref{eq:Dterm}), it is difficult to realize the vanishing $D$-terms 
for $h^{1,1}=1$. 
Hence, we demand $h^{1,1}\geq 2$, yielding three massive $U(1)$ gauge bosons 
except for $U(1)_Y$ and $U(1)_{B-L}$. 

One option to break the Pati-Salam symmetry is its spontaneous symmetry breaking in the presence of 
vector-like Higgs $(\bar{4}, 1,2,1)+{\rm c.c.}$. 
In this paper, we take into account another option, namely to introduce Wilson lines into the internal components of 
$SU(4)_C$ and $SU(2)_R$ on the quotient CY manifolds.

%%%%%%%%%%%%%%%%%%%%%%%%%%%%%%%%%%%%%%%%%%%%%%%%%%%%%%%%%%%%%%%%%%%%%%%%%%%%%%%%%%%%%%%%%%%%%%%%%%%
%%%%%%%%%%%%%%%%%%%%%%%%%%%%%%%%%%%%%%%%%%%%%%%%%%%%%%%%%%%%%%%%%%%%%%%%%%%%%%%%%%%%%%%%%%%%%%%%%%%
\section{Three-generation models on quotient complete intersection CY manifolds}
\label{sec:3}
We are now ready to searching for the three-generation models on CY manifolds with multiple line bundles 
satisfying all the consistency conditions. 
In particular, we concentrate on smooth, quotient CICYs 
${\cal \tilde{M}}={\cal M}/\Gamma$~\cite{Candelas:1987kf,Candelas:1987du}, where $\cal{M}$ is a simply connected CICY with a freely-acting discrete symmetry group $\Gamma$. 
CICYs defined in the ambient space $\mathbb{P}^{n_1}\times \cdots \times \mathbb{P}^{n_m}$ are characterized by the following $m\times R$ configuration matrix,
\begin{align}
\begin{matrix}
\mathbb{P}^{n_1}\\
\mathbb{P}^{n_2}\\
\vdots\\
\mathbb{P}^{n_m}\\
\end{matrix}
\begin{bmatrix}
q_1^1 & q_2^1 & \cdots & q_R^1\\
q_1^2 & q_2^2 & \cdots & q_R^2\\
\vdots & \vdots & \ddots & \vdots\\
q_1^m & q_2^m & \cdots & q_R^m\\
\end{bmatrix}
_{\chi}^{h^{1,1},h^{1,2}},
\end{align} 
where the subscript and superscripts denote the Euler number and Hodge numbers of CICYs respectively. 
When the homogeneous coordinates of $\mathbb{P}^{n_l}$ are represented as $x^l_\alpha$ 
with $l=1,\cdots, m$ and $\alpha=0,\cdots,n_l$, the positive integers $q^l_r$ $(r=1,\cdots, R)$ 
specify the multi-degree of $R$ homogeneous polynomials on $\mathbb{P}^{n_1}\times \cdots \times \mathbb{P}^{n_m}$. 
Then, the common zero locus of such $R$ polynomials corresponds to the defining equations of CICYs. 
The CY condition, namely a vanishing first Chern class of the tangent bundle, is realized by setting $\sum_{r=1}^Rq_r^l=n_l+1$, 
for all $l$ and 
the dimension of CICYs becomes $3$ under $\sum_{l=1}^m n_l-R=3$. 
Through out this paper, we focus on the favorable CICYs where the second cohomology of CY descends from that of 
the ambient space. 

Among them, some of CICYs admit a freely-acting discrete symmetry group $\Gamma$ classified in Ref.~\cite{Braun:2010vc}. 
Those quotient CICYs are of particular importance, since one can turn on Wilson lines because of the existence of the discrete symmetry group. Furthermore, $\Gamma$ reduces the number of complex structure moduli.\footnote{In general, the number of 
K\"ahler moduli on ${\cal M}$ and ${\cal \tilde{M}}={\cal M}/\Gamma$ is different from each other, but in this paper we focus on 
CICYs with $h^{1,1}({\cal M})=h^{1,1}({\cal \tilde{M}})$.} 
As discussed in Sec.~\ref{sec:2_4}, some of the line bundles satisfying all the conditions in Sec.~\ref{sec:2} 
should be correlated with one another. The gauge symmetry of the standard model in turn is enhanced to the Pati-Salam-like 
symmetry. We therefore introduce Wilson lines into the internal components of U(1)s on 
the quotient CICYs such that the Pati-Salam group is broken to the standard model one. 
It turns out that one cannot find realistic models on CICYs with $h^{1,1}=1,2,3,5$, since 
the underlying CY geometries and consistency conditions constrain the choice of background line bundles. 
In the next but one subsection~\ref{subsec:3_2}, 
we show the standard-like models on a concrete quotient CICY with $h^{1,1}({\cal \tilde{M}})=4$.

%%%%%%%%%%%%%%%%%%%%%%%%%%%%%%%%%%%%%%%%%%%%%%%%%%%%%%%%%%%%%%%%%%%%%%%%%%%%%%%%%%%%%%%%%%%%%%%%%%%
\subsection{Standard-like models via the Pati-Salam-like symmetry}
\label{subsec:3_1}
To break the Pati-Salam-like symmetry into the standard model one, 
we consider the quotient CICYs ${\cal \tilde{M}}={\cal M}/\Gamma$, where one can introduce 
Wilson lines into the internal components of $U(1)_{3,4,5}$ such that 
$SU(4)_C$ and $SU(2)_R\times U(1)^\prime$ are broken to $SU(3)_C\times U(1)_3$ and $U(1)_{4,5}$ respectively. 
(For more details, we refer to Ref.~\cite{Green:1987mn}.) 
When the discrete symmetry group $\Gamma$ consists of only Abelian discrete symmetries $\mathbb{Z}_n$, namely 
$\Gamma =\otimes_i \mathbb{Z}_{n_i}$, the representations of $\mathbb{Z}_{n_i}$ are represented as $e^{2\pi i p_i/n_i}$ 
for $p_i \in \{0,1,\cdots, n_i-1\}$. 
If the Wilson lines are in the center of $SU(4)_C$ and $SU(2)_R$, the Pati-Salam symmetry is unbroken. 
Hence, one of the Abelian symmetries should be different from $\mathbb{Z}_4$ and $\mathbb{Z}_2$. 
In such a case, the low-energy gauge group on the quotient CICYs ${\cal \tilde{M}}$ becomes
\begin{align}
SO(32)\rightarrow SU(3)_C\times SU(2)_L \times \Pi_{c=1}^5U(1)_c \times SO(16),
\end{align}
where three U(1)s except for $U(1)_Y$ and $U(1)_{B-L}$ become massive through the axionic couplings.

On quotient CICYs ${\cal \tilde{M}}$, the net-number of massless chiral supermultiplets is subject to the discrete 
symmetry group $\Gamma$. 
Indeed, the gauge bundles on ${\cal M}$ reduce to those on ${\cal \tilde{M}}$ 
if such bundles are equivariant bundles, and the cohomology associated with the matter field 
on ${\cal \tilde{M}}$ is described by the subspace of $H^1({\cal M}, V)$. Note that all line bundles 
have an equivariant structure for a single $\mathbb{Z}_n$, namely $\Gamma=\mathbb{Z}_n$. 
(For more details, we refer to Ref.~\cite{Anderson:2012yf}.) 
The number of generation on ${\cal M}$ is divided by the group order, $|\Gamma|$:  
\begin{align}
\chi({\cal \tilde{M}}, {\cal C}_p\otimes {\cal S})=\frac{\chi({\cal M}, {\cal C}_p)}{|\Gamma|},
\end{align}
where ${\cal S}$ is a certain representation of $\Gamma$. 
Note that the chiral asymmetry is independent of Wilson lines. 
%When the chiral asymmetry on ${\cal M}$ is smaller than unity, that is, $\chi({\cal \tilde{M}}, {\cal C}_p\otimes {\cal S}) <|\Gamma|$, it implies that the corresponding massless chiral modes are projected out on ${\cal M}$. 

Under these circumstances, we search for three generations of quarks and leptons 
such that the indices in Table~\ref{tab:3_1} become
\begin{align}
&
m_{(4,2,1,1)_{1,1,0}}+m_{(4,2,1,1)_{-1,1,0}}=\chi({\cal M}, L_1\otimes L_2)+\chi({\cal M}, L_1^{-1}\otimes L_2)=-3|\Gamma|,
\nonumber\\
&m_{(\bar{4},1,2,1)_{0,-1,-1}}+m_{(\bar{4},1,2,1)_{0,-1,1}}=
\chi({\cal M}, L_2^{-1}\otimes L_4^{-1})+\chi({\cal M}, L_2^{-1}\otimes L_4)=-3|\Gamma|,
%\nonumber\\
%&m_{d_R^c}=m_{d_{R_1}^c}+m_{d_{R_2}^c}+m_{d_{R_3}^{c\,4}}+m_{d_{R_3}^{c\,5}}
%=-\chi(L_2^{-1}\otimes L_4^{1})/|\Gamma|
%-\chi(L_2^{-1}\otimes L_5^{1})/|\Gamma|=3,
%\nonumber\\
%&m_{e_{R_1}^c}=m_{e_{R_1}^{c\,4}}+m_{e_{R_1}^{c\,5}}+m_{e_{R_2}^{c\,45}}
%=-\chi(L_2^{-1}\otimes L_4^1)/|\Gamma|
%-\chi(L_2^{-1}\otimes L_5^1)/|\Gamma|-\chi(L_4^{1}\otimes L_5^{1})/|\Gamma|=3,
\label{eq:3gen1}
\end{align}
within $-3|\Gamma| \leq m_{(4,2,1,1)_{\pm 1,1,0}}$, $m_{(\bar{4},1,2,1)_{0,-1,\pm 1}}\leq 0$ and indices of other massless states in Table~\ref{tab:3_1} are integrally quantized on ${\cal M}$ and ${\cal \tilde{M}}$. 
Now, we have used $c_1(L_3)=0$ and each index is calculated using the formula in 
Appendix~\ref{app}. 
Note that we have not taken into account the generation of Higgs and Higgsino fields which could be identified with $L_{3,4}^{4,5}$ and/or their conjugates in the latter analysis. 
We further require that the exotic modes such as $(4,1,1,16)_{0,-1,0}$, $(1,2,1,16)_{1,0,0}$ and $(1,1,2,16)_{0,0,1}$ in Table~\ref{tab:3_1} 
should not be chiral, namely 
\begin{align}
&\chi(L_1)=0,
\qquad 
\chi(L_2)=0,
\qquad 
\chi(L_4)=-\chi(L_5)=0. 
\label{eq:3gen2}
\end{align}
When the heterotic five-branes present in the system, 
there exist chiral fermions under the fundamental representations of $SO(32)\times Sp(2N_s)$, 
namely $(32, 2N_s)$. Here, $N_s$ denotes the number of heterotic five-branes which have the symplectic gauge degrees of freedom as confirmed in the S-dual type I superstring theory~\cite{Witten:1995gx}. 
Since these chiral fermions have $(4,1,1,16, 2N_s)_{0,-1,0}$, $(1,2,1,16, 2N_s)_{1,0,0}$ and $(1,1,2,16, 2N_s)_{0,0,1}$ 
representations of $ SU(4)_C\times SU(2)_L \times SU(2)_R \times \Pi_{c=1}^2U(1)_c \times U(1)^\prime \times SO(16) \times Sp(2N_s)$, 
the condition $(\ref{eq:3gen2})$ also ensures the absence of these chiral fermions. 
Here, the subscript indices label the $U(1)_1$, $U(1)_2$ and $U(1)^\prime$ charges.

To count the number of vector-like pairs for each representation, 
we have to calculate the dimensions of its corresponding cohomology $H^\ast ({\cal M}, {\cal C}_p)$ 
where ${\cal C}_p$ denotes the internal bundle. 
For our purpose, it is sufficient to consider the number of chiral modes to derive the particles in 
the standard-like models. Recall that the Higgs and higgsino fields will be identified with 
$L_{3,4}^{4,5}$ and/or their conjugates in the latter analysis. Although it is possible to break the Pati-Salam-like symmetry by 
the vector-like Higgs $(\bar{4}, 1,2,1)+{\rm c.c.}$, in this paper, we concentrate on Wilson lines to break the 
Pati-Salam-like symmetry, namely $SU(4)_C\rightarrow SU(3)_C\times U(1)_3$ and $SU(2)_R\times U(1)^\prime\rightarrow 
U(1)_4\times U(1)_5$. 
It can be achieved by the introduction of discrete Wilson lines on the quotient CICYs ${\cal \tilde{M}}={\cal M}/\Gamma$, 
where $\Gamma$ should be different from $Z_4$ and $Z_2$. 
However, in such a scenario, $U(1)_{B-L}$ gauge symmetry is still unbroken on the quotient CICYs. 
One of the possibilities to break $U(1)_{B-L}$ gauge symmetry is the 
non-zero vacuum expectation value of at least one of the right-handed sneutrino 
which carries the $U(1)_{B-L}$ charge not the $U(1)_Y$ charge. 
When the soft mass of the right-handed sneutrino is tachyonic at the TeV scale through renormalization group effects, 
the sneutrino develops a vacuum expectation value. 
Such a spontaneous $U(1)_{B-L}$ breaking is demonstrated 
in the B-L MSSM derived from the $E_8\times E_8$ heterotic string theory~\cite{Ambroso:2009jd,Ovrut:2012wg}. 
%Furthermore, the mass of the $Z^\prime$ boson becomes sufficiently larger than the current experimental bound. 
This $U(1)_{B-L}$ breaking scenario requires a deeper understanding of the mechanism of supersymmetry breaking 
and the stabilization of moduli fields. It will be the subject of future work.

%Those CY data is extracted from the toric data packaged in PALP~\cite{}. 

\begin{table}[htb]
\centering
  \begin{tabular}{|c|c|c|c|} \hline
    $\left(SU(4)_C\times SU(2)_L\times SU(2)_R\right.$ & $\left(SU(3)_C\times SU(2)_L\times SO(16)\right.$ 
    & & \\
    $\left.\times SO(16)\right)_{U(1)_1,U(1)_2,U(1)_4, U(1)_5}$ & $\left.\right)_{U(1)_1,U(1)_2,U(1)_3,U(1)_4,U(1)_5}$  
    & Matter &Index\\ \hline \hline
    $(4,2,1,1)_{1,1,0,0}$ & $(3,2,1)_{1,1,1,0,0}$ & $Q_1$ & $\chi({\cal M},  L_1\otimes L_2)/|\Gamma|$ 
  \\
     & $(1,2,1)_{1,1,-3,0,0}$ & $L_1$ &    
    \\ \hline
    $(4,2,1,1)_{-1,1,0,0}$ & $(3,2,1)_{-1,1,1,0,0}$ & $Q_2$ & $\chi({\cal M},  L_1^{-1}\otimes L_2)/|\Gamma|$  
    \\
    & $(1,2,1)_{-1,1,-3,0,0}$ & $L_2$ &   
    \\ \hline
    $(15,1,1,1)_{0,0,0,0}$ & $({\bar 3},1,1)_{0,0,-4,0,0}$ & $u_{R_1}^c$ & $\chi({\cal M}, {\cal O}_{\cal M})/|\Gamma|$  
    \\ \hline
    $(6,1,1,1)_{0,2,0,0}$ & $({\bar 3},1,1)_{0,2,2,0,0}$ & $d_{R_1}^c$ & $\chi({\cal M},  L_2^2)/|\Gamma|$  
    \\ 
     & $(3,1,1)_{0,2,-2,0,0}$ & $\bar{d}_{R_2}^c$ &   
    \\ \hline
     $(1,1,1,1)_{2,0,0,0}$ & $(1,1,1)_{2,0,0,0,0}$ &  $n_1$ & $\chi({\cal M},  L_1^{2})/|\Gamma|$  
        \\ \hline
    $(\bar{4}, 1,2,1)_{0,-1,-1,0}$ & $({\bar 3},1,1)_{0,-1,-1,-1,0}$ & $u_{R_2}^{c\,\,4}$ & $\chi({\cal M}, L_2^{-1} \otimes L_4^{-1})/|\Gamma|$   
    \\ 
       & $(1,1,1)_{0,-1,3,0,1}$ &  $e_{R_1}^{c\,\,5}$ &    
    \\ 
        & $(1,1,1)_{0,-1,3,-1,0}$ & $n_{2}^{c\,4}$ &    
    \\ 
    & $({\bar 3},1,1)_{0,-1,-1,0,1}$ & $d_{R_3}^{c\,\,5}$ &   
    \\ \hline
    $(\overline{4}, 1, 2, 1)_{0,-1,1,0}$  & $({\bar 3},1,1)_{0,-1,-1,1,0}$ & $d_{R_3}^{c\,\,4}$ & $\chi({\cal M}, L_2^{-1}\otimes L_4)/|\Gamma|$   
    \\ 
       &$(1,1,1)_{0,-1,3,1,0}$ &  $e_{R_1}^{c\,\,4}$ &    
    \\ 
        & $(1,1,1)_{0,-1,3,0,-1}$ & $n_{2}^{c\,5}$ &    
    \\ 
    & $({\bar 3},1,1)_{0,-1,-1,0,-1}$ & $u_{R_2}^{c\,\,5}$ &    
    \\ \hline
    $(1,2,2,1)_{1,0,-1,0}$  & $(1,2,1)_{1,0,0,-1,0}$ & $L_3^4$ & $\chi({\cal M},  L_1\otimes L_4^{-1})/|\Gamma|$  
    \\ 
      & $(1,2,1)_{1,0,0,0,1}$ & $\bar{L}_4^5$ &    
    \\ \hline 
    $(1,2,2,1)_{1,0,1,0}$  & $(1,2,1)_{1,0,0,0,-1}$ &  $L_3^5$ &  $\chi({\cal M},  L_1\otimes L_4)/|\Gamma|$  
    \\ 
      & $(1,2,1)_{1,0,0,1,0}$ & $\bar{L}_4^4$ &    
    \\ \hline    
    $(1,1,3,1)_{0,0,0,0}$ & $(1,1,1)_{0,0,0,1,1}$ & $e_{R_2}^{c\,\,45}$ & $\chi({\cal M}, {\cal O}_{\cal M})/|\Gamma|$   
    \\ 
     $(1,1,1,1)_{0,0,2,0}$ & $(1,1,1)_{0,0,0,1,-1}$ & $n_{3}^{c\,45}$ &  $\chi({\cal M}, L_4^{2})/|\Gamma|$   
    \\ \hline
    $(\bar{4},1,1,16)_{0,-1,0,0}$ & $({\bar 3},1,16)_{0,-1,-1,0,0}$ & $-$ & $\chi({\cal M}, L_2^{-1})/|\Gamma|$   
    \\
      & $(1,1,16)_{0,-1,3,0}$ & $-$ &    
    \\ \hline    
    $(1,2,1,16)_{1,0,0,0}$ & $(1,2,16)_{1,0,0,0,0}$ & $-$ & $\chi({\cal M}, L_1)/|\Gamma|$   
    \\  \hline
    $(1,1,2,16)_{0,0,1,0}$ & $(1,1,16)_{0,0,0,1,0}$ & $-$ & $\chi({\cal M}, L_4)/|\Gamma|$   
    \\ 
     & $(1,1,16)_{0,0,0,0,-1}$ & $-$ &    
    \\ \hline    
     $(1,1,1,120)_{0,0,0,0}$ & $(1,1,120)_{0,0,0,0,0}$ &  $-$ & $\chi({\cal M}, {\cal O}_{\cal M})/|\Gamma|$   
    \\ \hline
  \end{tabular}
  \caption{Massless spectrum and corresponding index. 
  In the first and second columns, the entries refer to the representations under $\left(SU(4)_C\times SU(2)_L\times SU(2)_R\times SO(16)\right)_{U(1)_1,U(1)_2,U(1)_4,U(1)_5}$ and $\left(SU(3)_C\times SU(2)_L\times SO(16)\right)_{U(1)_1,U(1)_2,U(1)_3,U(1)_4,U(1)_5}$ 
  and the subscript indices label the $U(1)$ charges. 
  Note that in the first column, two $U(1)_{4,5}$ charge vectors 
${\bm q}=(q_4, q_5)$ and ${\bm q}^\prime=(q_4^\prime, q_5^\prime)$ are identified if ${\bm q}-{\bm q}^\prime \in \mathbb{Z}\times (1,1)$ because of $L_5\simeq L_4^{-1}$.}
  \label{tab:3_1}
\end{table}

\clearpage
%%%%%%%%%%%%%%%%%%%%%%%%%%%%%%%%%%%%%%%%%%%%%%%%%%%%%%%%%%%%%%%%%%%%%%%%%%%%%%%%%%%%%%%%%%%%%%%%%%%%%
\subsection{Explicit three-generation models}
\label{subsec:3_2}
Let us search for the models passing all the requirements on the quotient CICYs within $2\leq h^{1,1}\leq 5$. 
%the supersymmetric condition is not trivially satisfied at the tree-level as shown in Eq.~(\ref{eq:Dterm}). 
For the 6 CICYs with $h^{1,1}({\cal M})=2$ and 12 CICYs with $h^{1,1}({\cal M})=3$, there are totally four quotient CICYs where one of the discrete symmetry is different from $\mathbb{Z}_2$ and $\mathbb{Z}_4$~\cite{Braun:2010vc}. 
They are characterized by the configuration matrices listed in App.~\ref{app:B}. 
However, one cannot find realistic models within the range $|m_a^i|\leq 8$. 

In a similar way, we next search for the models on ${\cal M}$ with $h^{1,1}({\cal M})=4$. 
Although there exist two possible CICYs different from $\mathbb{Z}_2$ and $\mathbb{Z}_4$ for the 19 CICYs with 
$h^{1,1}({\cal M})=4$, the following CICY, 
\begin{align}
\begin{matrix}
\mathbb{P}^2\\
\mathbb{P}^2\\
\mathbb{P}^2\\
\mathbb{P}^2\\
\end{matrix}
\begin{bmatrix}
1 & 1 & 1 & 0 & 0\\
1 & 1 & 0 & 1 & 0\\
1 & 1 & 0 & 0 & 1\\
0 & 0 & 1 & 1 & 1\\
\end{bmatrix}
^{4, 40}_{-72},
\label{eq:7247}
\end{align} 
leads to realistic models. It admits the $\mathbb{Z}_3$ and $\mathbb{Z}_3\times \mathbb{Z}_3$ 
freely-acting discrete symmetries. The nonvanishing intersection numbers and second Chern 
number of the tangent bundle expressed in terms of a basis $\tilde{w}^i$ on ${\cal M}$ are given by
\begin{align}
&d_{123}=6,
\qquad
d_{124}=d_{134}=d_{234}=5,
\qquad
d_{112}=d_{113}=d_{122}=d_{133}=d_{223}=d_{233}=3,
\nonumber\\
&d_{114}=d_{144}=d_{224}=d_{244}=d_{334}=d_{344}=2,
\end{align}
and 
\begin{align}
c_2(T{\cal M})=(36, 36, 36, 36),
\end{align}
respectively. 
Since one cannot find realistic models satisfying all the consistency conditions within the range $|m_a^i|\leq 4$ 
for the $\mathbb{Z}_3\times \mathbb{Z}_3$ case, we focus on the $\mathbb{Z}_3$ case where 
the number of complex structure moduli as well as the Euler number on the quotient CY manifold ${\cal {\tilde{M}}}={\cal M}/\mathbb{Z}_3$ reduce to be $h^{2,1}=16$ and $\chi=-24$. 
%We consider the line bundles of Eq.~(\ref{eq:ansatz}) 
%such that the K-theory condition and $U(1)_Y$ massless conditions are satisfied at the same time. 
%In addition, these first Chern numbers are further constrained by the anomaly cancellation condition in Eq.~(\ref{eq:anomalysec3}). 

When we search for the line bundles within the range $|m_a^i|\leq 4$ to 
realize Eqs.~(\ref{eq:3gen1}) and~(\ref{eq:3gen2}), namely the three generations of quarks and leptons 
and no chiral exotics on the quotient CICY ${\cal \tilde{M}}$, it turns out that the lists of the first Chern numbers 
in Tables~\ref{tab:3_2_1} and~\ref{tab:3_2_1_2} yield realistic models. 
In both cases with Tables~\ref{tab:3_2_1} and~\ref{tab:3_2_1_2}, the rank of the mass matrix of U(1)s in Eq.~(\ref{eq:Mai}) is $3$ 
and the remaining two $U(1)$ gauge symmetries are $U(1)_Y$ and $U(1)_{B-L}=U(1)_3/3$. 
The distinction between the spectrum of Tables~\ref{tab:3_2_1} and~\ref{tab:3_2_1_2} are the existence of Higgs fields 
which could be identified with $L_{3,4}^{4,5}$ and/or their conjugates. 
The indices of Higgs fields $L_{3,4}^{4,5}$ vanish for the choice of line bundles in Table~\ref{tab:3_2_1}, 
whereas Higgs fields appear in the case of Table~\ref{tab:3_2_1_2}. 
As commented before, the exotic modes such as $d_{R_1}$ and $d_{R_2}^c$ are the vector-like particles 
with respect to the standard model gauge group with an extra $U(1)_{B-L}=U(1)_3/3$ on the quotient CICY. 
Indeed, the net-number of chiral modes is the same because of $\chi(L_2^2)=-\chi(L_2^{-2})$. 
We expect that such modes could become massive due to the loop-effects for the anomalous $U(1)_2$ symmetry. 
Since the obtained spectrum illustrated in Table~\ref{tab:3_2_2} consists of the three generations of quarks and leptons, 
vector-like Higgs and extra vector-like particles with respect to $SU(3)_C\times SU(2)_L\times U(1)_Y\times U(1)_{B-L}$, 
this models are free of gauge and gravitational anomalies. 
%Note that we list the particles carrying MSSM quantum numbers in Table~\ref{tab:3_2_2}. 
For other examples in Table~\ref{tab:3_2_1_2}, the particle spectrum is similar to that in Table~\ref{tab:3_2_2}, 
but the flavor structure of Yukawa couplings could be different from one another. 
Indeed, for some of the line bundles 
listed in Table~\ref{tab:3_2_1_2}, the number of chiral generation is determined by only $Q_1(L_1)$ or $Q_2(L_2)$ 
in contrast to the case of Table~\ref{tab:3_2_2}.

Let us take a closer look at the possible Yukawa couplings among the standard model particles with an emphasis on the 
case of Table~\ref{tab:3_2_2}. 
In this model, we find that the gauge symmetries of the low-energy effective action allow 
for the following Yukawa couplings of quarks and leptons,
\begin{align}
&(Q_1, \bar{L}_3^5, u_{R_2}^{c\,\,5}),\qquad (Q_2, \bar{L}_4^5, u_{R_2}^{c\,\,5}),\qquad
(Q_1, L_4^4, d_{R_3}^{c\,\,4}), \qquad (Q_2, L_3^4, d_{R_3}^{c\,\,4}),
\nonumber\\
&(L_1, L_4^4, e_{R_1}^{c\,\,4}),\qquad (L_2, L_3^4, e_{R_1}^{c\,\,4}),\qquad
(L_1, \bar{L}_3^5, n_{2}^{c\,\,5}),\qquad (L_2, \bar{L}_4^5, n_{2}^{c\,\,5}),\qquad
\end{align}
by identifying the up-and down-type Higgs doublets as $\bar{L}_{3, 4}^5$ and $L_{3, 4}^4$. 
Here, $\bar{L}_4^5$ stands for the conjugate representation of $L_4^5$ and those Higgs fields are 
vector-like particles under the standard model gauge group with an extra $U(1)_{B-L}$. 
%it will become massive though the couplings such as $\bar{n}_1n_3^{c\,45}L_3^4\bar{L}_4^5$. 
Three right-handed neutrinos are identified with $n_2^{5}$. 
Although those couplings are allowed by the gauge symmetries, 
there exist topological selection rules for their Yukawa couplings 
as discussed in Refs.~\cite{Candelas:1987se,Anderson:2009ge,Blesneag:2015pvz}. 
We leave the detailed study of the Yukawa couplings for a future work. 
With the help of $U(1)_{B-L}$ and other anomalous $U(1)$ symmetries, 
R-parity violating and the dimension-four and -five proton decay operators such as $QQQL$ and $u^cu^cd^ce^c$ are forbidden. 
As stated already, we will also postpone the detailed study of $U(1)_{B-L}$ breaking for a future analysis. 

For the 23 CICYs with $h^{1,1}=5$, we have searched for the three-generation models 
and among them, there are four possible candidates where 
one of the discrete symmetry group is different from $\mathbb{Z}_2$ and $\mathbb{Z}_4$~\cite{Braun:2010vc}. 
However, it turns out that the line bundle background does not pass all the requirements within the range $|m_a^i|\leq 4$.

\begin{table}[htb]
\centering
  \begin{tabular}{|c|c|c|c|c|} \hline
    $(m_1^1, m_1^2,m_1^3,m_1^4)$ & $(m_2^1, m_2^2 ,m_2^3,m_2^4)$ & $(m_4^1, m_4^2, m_4^3,m_4^4)$ & \# of Five-branes\\ \hline 
    $(\mp 2,0, \pm 1, 0)$ & $(0, 1,-2,0)$  & $(\mp 1,\pm 1,0, 0)$ & $(6, 24, 12, 0)$\\ \hline
   % $(\mp 1, \pm 1,0, 0)$ & $(0, -1, 2, 0)$  & $(\mp 2,0, \pm 1, 0)$ & $(6, 24, 12, 0)$\\ \hline
  \end{tabular}
  \caption{Number of heterotic five-branes in the basis $(w_1, w_2,w_3,w_4)$ and the first Chern numbers $c_1^i(L_a)=m_a^i$ 
  in agreement with all the requirements, 
  where other line bundles are constrained as $c_1(L_5)=-c_1(L_4)$ and $c_1(L_3)=0$. We omit the other sign flipping and 
  possible permutations among the $m_a^i$. The list of this table gives rise to the vanishing indices of Higgs fields $L_{3,4}^{4,5}$.}
  \label{tab:3_2_1}
\end{table}

\begin{table}[htb]
\centering
  \begin{tabular}{|c|c|c|c|c|} \hline
    $(m_1^1, m_1^2,m_1^3,m_1^4)$ & $(m_2^1, m_2^2 ,m_2^3,m_2^4)$ & $(m_4^1, m_4^2, m_4^3,m_4^4)$ & \# of Five-branes\\ \hline 
    $(\mp 3,0, 0, \pm 3)$ & $(1, 0,-1,2)$  & $(0, \mp 1,\pm 1, \pm 4)$ & $(30, 48, 30, 0)$\\ \hline
    %$(0, \mp 1,\pm 1, \pm 4)$ & $(-1, -1, 2, 0)$  & $(\mp 2,0, \pm 1, 0)$ & $(30, 48, 30, 0)$\\ \hline
    $(\mp 1,0, \pm 1, \mp 3)$ & $(0, 1,-1,-2)$  & $(\mp 1,\pm 1,0, \pm 1)$ & $(42, 75, 51, 12)$\\ \hline
    %$(\mp 1, \pm 1,0, \pm 1)$ & $(0, -1,1,2)$  & $(\mp 1,0, \pm 1, \mp 3)$ & $(42, 75, 51, 12)$\\ \hline
    $(\mp 1, 0, \pm 1, \mp 3)$ & $(1, -1, 0,-1)$  & $(0, \mp 1, \pm 1, \mp 1)$ & $(39, 27, 66, 12)$\\ \hline
    %$(0, \mp 1, \pm 1, \mp 1)$ & $(-1, 1, 0,1)$  & $(\mp 1, 0, \pm 1, \mp 3)$ & $(39, 27, 66, 12)$\\ \hline
    $(\mp 1, 0, \pm 1, \mp 2)$ & $(0, 1, -1,-1)$  & $(\mp 1, \pm 1, 0, 0)$ & $(18, 45, 33, 12)$\\ \hline
    %$(\mp 1, \pm 1, 0, 0)$ & $(0, -1, 1,1)$  & $(\mp 1, 0, \pm 1, \mp 2)$ & $(18, 45, 33, 12)$\\ \hline
  \end{tabular}
  \caption{Number of heterotic five-branes in the basis $(w_1, w_2,w_3,w_4)$ and the first Chern numbers $c_1^i(L_a)=m_a^i$ 
  in agreement with all the requirements, 
  where other line bundles are constrained as $c_1(L_5)=-c_1(L_4)$ and $c_1(L_3)=0$. We omit the other sign flipping and 
  possible permutations among the $m_a^i$. One of the chiral spectrum is illustrated in Table~\ref{tab:3_2_2}.}
  \label{tab:3_2_1_2}
\end{table}

\begin{table}[htb]
\centering
  \begin{tabular}{|c|c|c|c|c|} \hline
    Matter & Repr. & Index & Multiplicities & Total\\ \hline \hline
    $Q_1$ & $(3,2,1)_{1,1,1,0,0}$ & $\chi({\cal M},  L_1\otimes L_2)/|\Gamma|$ 
    & -1 &\\
    $Q_2$ & $(3,2,1)_{-1,1,1,0,0}$ & $\chi({\cal M}, L_1^{-1}\otimes L_2)/|\Gamma|$   
    & -2 & -3\\ \hline
    $L_1$ & $(1,2,1)_{1,1,-3,0,0}$ & $\chi({\cal M},  L_1\otimes L_2)/|\Gamma|$  
    & -1 &\\
    $L_2$ & $(1,2,1)_{-1,1,-3,0,0}$ & $\chi({\cal M}, L_1^{-1}\otimes L_2)/|\Gamma|$ 
    & -2 &\\
    $L_3^4$ & $(1,2,1)_{1,0,0,-1,0}$ & $\chi({\cal M},  L_1\otimes L_4^{-1})/|\Gamma|$  
    & -4 &\\
    $L_3^5$ & $(1,2,1)_{1,0,0,0,-1}$ & $\chi({\cal M},  L_1\otimes L_4)/|\Gamma|$  
    & -28 &\\
    $L_4^4$ & $(1,2,1)_{-1,0,0,-1,0}$ & $\chi({\cal M}, L_1^{-1}\otimes L_4^{-1})/|\Gamma|$   
    & 28 &\\
    $L_4^5$ & $(1,2,1)_{-1,0,0,0,-1}$ & $\chi({\cal M}, L_1^{-1}\otimes L_4)/|\Gamma|$   
    & 4 & -3\\ \hline
    $u_{R_1}^c$ & $({\bar 3},1,1)_{0,0,-4,0,0}$ & $\chi({\cal M},  {\cal O}_{\cal M})/|\Gamma|$  
    & 0 &\\
    $u_{R_2}^{c\,\,4}$ & $({\bar 3},1,1)_{0,-1,-1,-1,0}$ & $\chi({\cal M}, L_2^{-1}\otimes L_4^{-1})/|\Gamma|$   
    & 0 &\\
    $u_{R_2}^{c\,\,5}$ & $({\bar 3},1,1)_{0,-1,-1,0,-1}$ & $\chi({\cal M}, L_2^{-1} \otimes L_4)/|\Gamma|$   
    & -3 & -3 \\ \hline
    $d_{R_1}^c$ & $({\bar 3},1,1)_{0,2,2,0,0}$ & $\chi({\cal M},  L_2^2 )/|\Gamma|$  
    & -12 &\\
    $d_{R_2}^c$ & $({\bar 3},1,1)_{0,-2,2,0,0}$ & $\chi({\cal M},  L_2^{-2})/|\Gamma|$  
    & 12 &\\
    $d_{R_3}^{c\,\,4}$ & $({\bar 3},1,1)_{0,-1,-1,1,0}$ & $\chi({\cal M}, L_2^{-1} \otimes L_4)/|\Gamma|$   
    & -3 &\\
    $d_{R_3}^{c\,\,5}$ & $({\bar 3},1,1)_{0,-1,-1,0,1}$ & $\chi({\cal M}, L_2^{-1} \otimes L_4^{-1})/|\Gamma|$  
    & 0 & -3\\ \hline
    $e_{R_1}^{c\,\,4}$ & $(1,1,1)_{0,-1,3,1,0}$ & $\chi({\cal M}, L_2^{-1} \otimes L_4)/|\Gamma|$   
    & -3 &\\
    $e_{R_1}^{c\,\,5}$ & $(1,1,1)_{0,-1,3,0,1}$ & $\chi({\cal M}, L_2^{-1} \otimes L_4^{-1})/|\Gamma|$   
    & 0 &\\
    $e_{R_2}^{c\,\,45}$ & $(1,1,1)_{0,0,0,1,1}$ & $\chi({\cal M}, {\cal O}_{\cal M})/|\Gamma|$   
    & 0 & -3\\ \hline
    $n_1$ & $(1,1,1)_{2,0,0,0,0}$ & $\chi({\cal M},  L_1^{2})/|\Gamma|$  
    & 0 & 0\\
    $n_{2}^{c\,4}$ & $(1,1,1)_{0,-1,3,-1,0}$ & $\chi({\cal M}, L_2^{-1} \otimes L_4^{-1})/|\Gamma|$   
    & 0 & 0 \\
    $n_{2}^{c\,5}$ & $(1,1,1)_{0,-1,3,0,-1}$ & $\chi({\cal M}, L_2^{-1} \otimes L_4)/|\Gamma|$   
    & -3 & -3\\
    $n_{3}^{c\,45}$ & $(1,1,1)_{0,0,0,1,-1}$ & $\chi({\cal M}, L_4^{2})/|\Gamma|$   
    & -24 & -24\\ \hline
    $-$ & $(1,2,16)_{1,0,0,0,0}$ & $\chi({\cal M}, L_1)/|\Gamma|$   
    & 0 & 0\\
    $-$ & $({\bar 3},1,16)_{0,-1,-1,0,0}$ & $\chi({\cal M}, L_2^{-1} )/|\Gamma|$   
    & 0 & 0\\
    $-$ & $(1,1,16)_{0,-1,3,0,0}$ & $\chi({\cal M}, L_2^{-1})/|\Gamma|$   
    & 0 & 0\\
    $-$ & $(1,1,16)_{0,0,0,1,0}$ & $\chi({\cal M}, L_4)/|\Gamma|$   
    & 0 & 0\\
    $-$ & $(1,1,16)_{0,0,0,0,-1}$ & $\chi({\cal M}, L_4)/|\Gamma|$   
    & 0 & 0\\ \hline
    $-$ & $(1,1,120)_{0,0,0,0,0}$ & $\chi({\cal M}, {\cal O}_{\cal M})/|\Gamma|$   
    & 0 & 0\\ \hline
  \end{tabular}
  \caption{Supersymmetric spectrum in visible and hidden sectors for the line bundles 
   $(m_1^1, m_1^2,m_1^3,m_1^4)=(- 3,0, 0, 3)$, $(m_2^1, m_2^2 ,m_2^3,m_2^4)=(1, 0,-1,2)$ and 
   $(m_4^1, m_4^2, m_4^3,m_4^4)=(0, -1, 1, 4)$ in Table~\ref{tab:3_2_1_2}. 
  In the second column, the entries refer to the representations under $SU(3)_C$, $SU(2)_L$ and $SO(16)$ 
  and the subscript indices label the $U(1)$ charges.}
  \label{tab:3_2_2}
\end{table}

%%%%%%%%%%%%%%%%%%%%%%%%%%%%%%%%%%%%%%%%%%%%%%%%%%%%%%%%%%%%%%%%%%%%%%%%%%%%%%%%%%%%%%%%%%%%%%%%%%%%%%%%
\clearpage
\section{Conclusion}
\label{sec:con}
We have searched for the three-generation standard-like models from the $SO(32)$ heterotic 
string theory on quotient complete intersection Calabi-Yau manifolds with line bundles. 
The line bundles with $U(1)$ structure groups, in contrast to non-abelian bundles, 
are promising tool to realize standard-like models not only in 
the $E_8\times E_8$ heterotic string, but also the $SO(32)$ one. 
Such stable line bundles lead to the three chiral generations of quarks and leptons.

In this paper, we have presented that the Wilson lines and line bundles satisfying the supersymmetric conditions, 
the $U(1)_Y$ masslessness conditions and K-theory condition yield the standard-like model 
gauge group $SU(3)_C\times SU(2)_L\times U(1)_Y\times U(1)_{B-L}$ via a 
Pati-Salam-like symmetry. 
Other U(1) gauge bosons become massive through the Green-Schwarz mechanism. 
The obtained spectrum contains the three generations of quarks and leptons, 
vector-like Higgs and extra vector-like particles with respect to $SU(3)_C\times SU(2)_L\times U(1)_Y\times U(1)_{B-L}$. 
Although the Yukawa couplings are allowed by the gauge symmetries of the low-energy 
effective action, we leave the derivation of selection rules for the holomorphic Yukawa couplings on CICYs~\cite{Anderson:2009ge,Blesneag:2015pvz,Blesneag:2016yag} for a future work. 

Motivated by S- and T-dual models of intersecting D-branes in type IIA string theory, 
we have focused on the Pati-Salam-like gauge group to derive the standard 
model one. 
It would be interesting to check whether $SO(32)$ gauge group is 
directly broken to the standard model gauge group by a hypercharge flux, 
although there exists a no-go theorem on the existence of a standard model spectrum consistent with the 
hypercharge flux breaking and gauge coupling unification in the $E_8\times E_8$ heterotic string~\cite{Anderson:2014hia}. 
The gauge couplings of non-abelian gauge groups embedded in the $SO(32)$ heterotic string are non-universal 
in contrast to those in the $E_8\times E_8$ heterotic string~\cite{Blumenhagen:2005ga,Blumenhagen:2005pm}. 
Such non-universal gauge couplings have the potential to explain the differences between the gauge couplings of $SU(3)_C$ 
and $SU(2)_L$ at the string scale as demonstrated in the toroidal background~\cite{Abe:2015xua}.

\section*{Acknowledgments}
  We would like to thank H.~Abe, T.~Kobayashi, Y.~Takano, T.~H.~Tatsuishi and A.~Lukas for useful discussions and comments. 
  We also would like to thank the referee to improve the paper and 
  the organizing committee of the conference of ``String Phenomenology 2017'' 
  for their hospitality. 
  H.~O. was supported in part by Grant-in-Aid for Young Scientists (B) (No.~17K14303) 
  from Japan Society for the Promotion of Science.

\appendix 

\section{Net-number of chiral fields}
\label{app}
In this appendix, we present the index formula to calculate the net-number of chiral fields in Table~\ref{tab:3_1},
\begin{align}
\chi(L_a^p)
&=\int_{\cal M} \biggl[{\rm ch}_3(L_a^p)+\frac{c_2(T{\cal M})c_1(L_a^p)}{12}\biggl]
%\nonumber\\
%&=\int_{\cal M} \biggl[\frac{c_1^3(L_a)}{6}+\frac{c_2(T{\cal M})c_1(L_a)}{12}\biggl]
=d_{ijk}\biggl[\frac{p^3}{6}m_a^im_a^jm_a^k +\frac{p}{12}m_a^i c_2^{jk}(T{\cal M})\biggl],
\nonumber\\
\chi(L_a^p\otimes L_b^q)
&=\chi(L_a^p) +\chi(L_b^q)+\int_{\cal M} \biggl[ c_1(L_a^p){\rm ch}_2(L_b^q)+{\rm ch}_2(L_a^p)c_1(L_b^q)\biggl]
%\nonumber\\
%&=\chi(L_a) +\chi(L_b)+\frac{c_1(L_a)c_1^2(L_b)}{2}+\frac{c_1^2(L_a)c_1(L_b)}{2}
\nonumber\\
&=\chi(L_a^p) +\chi(L_b^q)+\frac{1}{2}d_{ijk}\biggl[pq^2 m_a^i m_b^j m_b^k +p^2q m_a^i m_a^j m_b^k\biggl],
%\nonumber\\
%\chi(L_a^p\otimes L_b^q \otimes L_c^r)
%&=\chi(L_a^p) +\chi(L_b^q \otimes L_c^r)+\int_{\cal M} \biggl[
%c_1(L_a^p){\rm ch}_2(L_b^q\otimes L_c^r)+{\rm ch}_2(L_a^p)c_1(L_b^q \otimes L_c^r)\biggl]
%\nonumber\\
%&=\chi(L_a) +\chi(L_b) +\chi(L_c)+\frac{c_1(L_b)c_1^2(L_c)}{2}+\frac{c_1^2(L_b)c_1(L_c)}{2}
%\nonumber\\
%&+c_1(L_a)\biggl[ {\rm ch}_2(L_b) +c_1(L_b)c_1(L_c)+ {\rm ch}_2(L_c)\biggl]+ 
% {\rm ch}_2(L_a)\biggl[ c_1(L_b)+c_1(L_c)\biggl]
% \nonumber\\
%&=\chi(L_a^p) +\chi(L_b^q) +\chi(L_c^r)+\int_{\cal M}\Biggl[c_1(L_a^p)c_1(L_b^q)c_1(L_c^r)
%+\frac{c_1^2(L_b^q)c_1(L_c^r)}{2}+
%\nonumber\\
%&\frac{c_1(L_b^q)c_1^2(L_c^r)+c_1(L_a^p)\left(c_1^2(L_b^q) +c_1^2(L_c^r)\right)
%+c_1^2(L_a^p)\left(c_1(L_b^q) +c_1(L_c^r)\right)}{2}
%\Biggl],
\end{align}
where $c_1^i(L_a)=m_a^i$ and $p,q,r$ are the integers.

\section{List of CICYs to be applicable for the Pati-Salam models}
\label{app:B}
To break the Pati-Salam-like symmetry by the inclusion of Wilson lines, 
the freely-acting discrete symmetry groups of CICYs should be different from $\mathbb{Z}_4$ and $\mathbb{Z}_2$. 
We list such topologically distinguishable CICYs for $2 \leq h^{1,1}\leq 5$ in Table~\ref{tab:app}~\cite{Braun:2010vc}, 
where $c_2(T{\cal M})$ denotes the second Chern number in the basis $\hat{w}^i$.  

\clearpage
\renewcommand{\arraystretch}{1.2}
\begin{table}[htbp]
\begin{center}
  \begin{minipage}[c]{15cm}
%\begin{longtable}{|c|c|l|c|}\hline
  \begin{tabular}{|c|c|c|c|} \hline
   $(h^{1,1}, h^{2,1})$ &  Configuration matrix & $c_2(T{\cal M})$ & $\Gamma$  \\
 \hline\hline
 (2, 52) & %-100 & 
\footnotesize$
\begin{matrix}
\mathbb{P}^4\\
\mathbb{P}^4\\
\end{matrix}
\begin{bmatrix}
1 & 1 & 1 & 1 & 1\\
1 & 1 & 1 & 1 & 1\\
\end{bmatrix}
$ & (24,24) & $Z_5$ \\ \hline
 (2, 56) & %-108 & 
\footnotesize$
\begin{matrix}
\mathbb{P}^2\\
\mathbb{P}^5\\
\end{matrix}
\begin{bmatrix}
 0 & 1 & 1 & 1\\
3 & 1 & 1 & 1\\
\end{bmatrix}
$ & (24,24,24,24,24) & $Z_3\times Z_3$ \\ \hline
 (2, 83) & %-162 & 
\footnotesize$
\begin{matrix}
\mathbb{P}^2\\
\mathbb{P}^2\\
\end{matrix}
\begin{bmatrix}
3\\
3\\
\end{bmatrix}
$ & (24,24,24,24,24) & $Z_3\times Z_3$ \\ \hline
 (3, 39) & %-72 & 
\footnotesize$
\begin{matrix}
\mathbb{P}^2\\
\mathbb{P}^2\\
\mathbb{P}^5\\
\end{matrix}
\begin{bmatrix}
1 & 1 & 1 & 0 & 0 & 0 \\
0 & 0 & 0 & 1 & 1 & 1 \\
1 & 1 & 1 & 1 & 1 & 1 \\
\end{bmatrix}
$ & (36, 36, 54) &  $Z_3$, $Z_3\times Z_3$ \\ \hline
 (3, 48) & %-90 & 
\footnotesize$
\begin{matrix}
\mathbb{P}^2\\
\mathbb{P}^2\\
\mathbb{P}^2\\
\end{matrix}
\begin{bmatrix}
1 & 1 & 1 \\
1 & 1 & 1 \\
1 & 1 & 1 \\
\end{bmatrix}
$ & (36, 36, 36) & $Z_3$, $Z_3\times Z_3$ \\ \hline
 (4, 68) & %-128 & 
\footnotesize$
\begin{matrix}
\mathbb{P}^1\\
\mathbb{P}^1\\
\mathbb{P}^1\\
\mathbb{P}^1\\
\end{matrix}
\begin{bmatrix}
2\\
2\\
2\\
2\\
\end{bmatrix}
$ & (24,24,24,24) & $Z_8$\footnote{Although this tetra-quadric has other 
discrete symmetries $Z_4$, $Z_2\times Z_2$, $Z_4\times Z_2$, $Q_8$, $Z_4\times Z_4$, $Z_4\rtimes Z_4$, 
$Z_8\times Z_2$, $Z_8\rtimes Z_2$, $Z_2\times Q_8$, we focus on 
the $Z_8$ case.}\\ \hline
 (4, 40) & %-72 & 
\footnotesize$
\begin{matrix}
\mathbb{P}^2\\
\mathbb{P}^2\\
\mathbb{P}^2\\
\mathbb{P}^2\\
\end{matrix}
\begin{bmatrix}
1 & 1 & 1 & 0 & 0\\
1 & 1 & 0 & 1 & 0\\
1 & 1 & 0 & 0 & 1\\
0 & 0 & 1 & 1 & 1\\
\end{bmatrix}
$ & (36, 36, 36, 36) & $Z_3$, $Z_3\times Z_3$ \\ \hline
 (5, 32) & %-54 & 
\footnotesize$
\begin{matrix}
\mathbb{P}^2\\
\mathbb{P}^2\\
\mathbb{P}^2\\
\mathbb{P}^2\\
\mathbb{P}^2\\
\end{matrix}
\!\left[
\begin{matrix}
1 & 1 & 1 & 0 & 0 & 0 & 0 \\
1 & 0 & 0 & 1 & 1 & 0 & 0 \\
1 & 0 & 0 & 0 & 0 & 1 & 1 \\
0 & 1 & 0 & 1 & 0 & 1 & 0 \\
0 & 0 & 1 & 0 & 1 & 0 & 1 \\
\end{matrix}
\right]
$ & (36,36,36,36,36)    &  $Z_3$ \\ \hline
 (5, 37) & %-64 & 
\footnotesize$
\begin{matrix}
\mathbb{P}^1\\
\mathbb{P}^1\\
\mathbb{P}^1\\
\mathbb{P}^1\\
\mathbb{P}^3\\
\end{matrix}
\!\left[
\begin{matrix}
0 & 0 & 0 & 2 \\
0 & 2 & 0 & 0 \\
0 & 0 & 2 & 0 \\
2 & 0 & 0 & 0 \\
1 & 1 & 1 & 1 \\
\end{matrix}
\right]
$ & (24,24,24,24,64)    &  $Z_8$ \\ \hline
 (5, 45) & %-80 & 
\footnotesize$
\begin{matrix}
\mathbb{P}^1\\
\mathbb{P}^1\\
\mathbb{P}^1\\
\mathbb{P}^1\\
\mathbb{P}^1\\
\end{matrix}
\!\left[
\begin{matrix}
1 & 1  \\
1 & 1  \\
1 & 1  \\
1 & 1  \\
1 & 1  \\
\end{matrix}
\right]
$ & (24,24,24,24,24)    &  $Z_5$, $Z_{10}$ \\ \hline
%\end{longtable}
  \end{tabular}
   \caption{List of favorable CICYs to be applicable for the Pati-Salam models.}
   \label{tab:app}
\end{minipage} 
\end{center}
 \end{table}

%\begin{longtable} {|c|c|}%{rllld{-2}r}
 %\caption{元素の名前と性質．}
 %\label{tab:elements_list}
 %\\
 %------ 最初のページの表の最上部 ----
%\hline
%\centering
% Configuration matrix & $\Gamma$  \\
% \hline\hline
% \endfirsthead
 %----------------------------------------------------------------
% \end{longtable}


\clearpage

\begin{thebibliography}{99}
%\cite{Gross:1985fr}
\bibitem{Gross:1985fr}
  D.~J.~Gross, J.~A.~Harvey, E.~J.~Martinec and R.~Rohm,
  %``Heterotic String Theory. 1. The Free Heterotic String,''
  Nucl.\ Phys.\ B {\bf 256} (1985) 253.  


  %\cite{Gross:1985rr}
\bibitem{Gross:1985rr}
  D.~J.~Gross, J.~A.~Harvey, E.~J.~Martinec and R.~Rohm,
  %``Heterotic String Theory. 2. The Interacting Heterotic String,''
  Nucl.\ Phys.\ B {\bf 267} (1986) 75.



%\cite{Candelas:1985en}
\bibitem{Candelas:1985en}
  P.~Candelas, G.~T.~Horowitz, A.~Strominger and E.~Witten,
  %``Vacuum Configurations for Superstrings,''
  Nucl.\ Phys.\ B {\bf 258} (1985) 46.  

%\cite{Greene:1986bm}
\bibitem{Greene:1986bm}
  B.~R.~Greene, K.~H.~Kirklin, P.~J.~Miron and G.~G.~Ross,
  %``A Three Generation Superstring Model. 1. Compactification and Discrete Symmetries,''
  Nucl.\ Phys.\ B {\bf 278} (1986) 667.  

  %\cite{Greene:1986jb}
\bibitem{Greene:1986jb}
  B.~R.~Greene, K.~H.~Kirklin, P.~J.~Miron and G.~G.~Ross,
  %``A Three Generation Superstring Model. 2. Symmetry Breaking and the Low-Energy Theory,''
  Nucl.\ Phys.\ B {\bf 292} (1987) 606. 
  
%\cite{Witten:1985bz}
\bibitem{Witten:1985bz}
  E.~Witten,
  %``New Issues in Manifolds of SU(3) Holonomy,''
  Nucl.\ Phys.\ B {\bf 268} (1986) 79.  

  
    
%\cite{Donagi:2000zf}
\bibitem{Donagi:2000zf}
  R.~Donagi, B.~A.~Ovrut, T.~Pantev and D.~Waldram,
  %``Standard model bundles on nonsimply connected Calabi-Yau threefolds,''
  JHEP {\bf 0108} (2001) 053
  [hep-th/0008008].

%\cite{Andreas:1999ty}
\bibitem{Andreas:1999ty}
  B.~Andreas, G.~Curio and A.~Klemm,
  %``Towards the Standard Model spectrum from elliptic Calabi-Yau,''
  Int.\ J.\ Mod.\ Phys.\ A {\bf 19} (2004) 1987
  [hep-th/9903052].

%\cite{Braun:2005ux}
\bibitem{Braun:2005ux}
  V.~Braun, Y.~H.~He, B.~A.~Ovrut and T.~Pantev,
  %``A Heterotic standard model,''
  Phys.\ Lett.\ B {\bf 618} (2005) 252
  [hep-th/0501070].
  
  
    %\cite{Blumenhagen:2005ga}
\bibitem{Blumenhagen:2005ga}
  R.~Blumenhagen, G.~Honecker and T.~Weigand,
  %``Loop-corrected compactifications of the heterotic string with line bundles,''
  JHEP {\bf 0506} (2005) 020
  [hep-th/0504232]. 

    %\cite{Blumenhagen:2005pm}
\bibitem{Blumenhagen:2005pm}
  R.~Blumenhagen, G.~Honecker and T.~Weigand,
  %``Supersymmetric (non-)Abelian bundles in the Type I and SO(32) heterotic string,''
  JHEP {\bf 0508} (2005) 009
  [hep-th/0507041].  
  

%\cite{Bouchard:2005ag}
\bibitem{Bouchard:2005ag}
  V.~Bouchard and R.~Donagi,
  %``An SU(5) heterotic standard model,''
  Phys.\ Lett.\ B {\bf 633} (2006) 783
  [hep-th/0512149].

    
  
%\cite{Narain:1986qm}
%\bibitem{Narain:1986qm}
%  K.~S.~Narain, M.~H.~Sarmadi and C.~Vafa,
%  %``Asymmetric Orbifolds,''
%  Nucl.\ Phys.\ B {\bf 288} (1987) 551;
%\cite{Ibanez:1987pj}
%\bibitem{Ibanez:1987pj}
%  L.~E.~Ibanez, J.~Mas, H.~P.~Nilles and F.~Quevedo,
%  %``Heterotic Strings in Symmetric and Asymmetric Orbifold Backgrounds,''
%  Nucl.\ Phys.\ B {\bf 301} (1988) 157.

%\cite{Anderson:2011ns}
\bibitem{Anderson:2011ns}
  L.~B.~Anderson, J.~Gray, A.~Lukas and E.~Palti,
  %``Two Hundred Heterotic Standard Models on Smooth Calabi-Yau Threefolds,''
  Phys.\ Rev.\ D {\bf 84} (2011) 106005
  %doi:10.1103/PhysRevD.84.106005
  [arXiv:1106.4804 [hep-th]].

  %\cite{Anderson:2012yf}
\bibitem{Anderson:2012yf}
  L.~B.~Anderson, J.~Gray, A.~Lukas and E.~Palti,
  %``Heterotic Line Bundle Standard Models,''
  JHEP {\bf 1206} (2012) 113
  %doi:10.1007/JHEP06(2012)113
  [arXiv:1202.1757 [hep-th]].
  

%\cite{Witten:1984dg}
\bibitem{Witten:1984dg}
  E.~Witten,
  %``Some Properties of O(32) Superstrings,''
  Phys.\ Lett.\ B {\bf 149} (1984) 351.
  %%CITATION = PHLTA,B149,351;%%
  %577 citations counted in INSPIRE as of 17 May 2014


%%\cite{Ibanez:2012zz}
%\bibitem{Ibanez:2012zz} 
%  L.~E.~Ibanez and A.~M.~Uranga,
%  %``String theory and particle physics: An introduction to string phenomenology,''
%  Cambridge, UK: Univ. Pr. (2012) 673 p


 

%\cite{Giedt:2003an}
\bibitem{Giedt:2003an}
  J.~Giedt,
  %``Z(3) orbifolds of the SO(32) heterotic string. 1. Wilson line embeddings,''
  Nucl.\ Phys.\ B {\bf 671} (2003) 133
%  doi:10.1016/j.nuclphysb.2003.08.031
  [hep-th/0301232].  
  
    %\cite{Choi:2004wn}
\bibitem{Choi:2004wn}
  K.~S.~Choi, S.~Groot Nibbelink and M.~Trapletti,
  %``Heterotic SO(32) model building in four dimensions,''
  JHEP {\bf 0412} (2004) 063
  %doi:10.1088/1126-6708/2004/12/063
  [hep-th/0410232].
  
  
  %\cite{Nilles:2006np}
\bibitem{Nilles:2006np}
  H.~P.~Nilles, S.~Ramos-Sanchez, P.~K.~S.~Vaudrevange and A.~Wingerter,
  %``Exploring the SO(32) heterotic string,''
  JHEP {\bf 0604} (2006) 050
  [hep-th/0603086]. 
  
  %\cite{RamosSanchez:2008tn}
\bibitem{RamosSanchez:2008tn}
  S.~Ramos-Sanchez,
  %``Towards Low Energy Physics from the Heterotic String,''
  Fortsch.\ Phys.\  {\bf 10} (2009) 907
  [arXiv:0812.3560 [hep-th]].


%\cite{Abe:2015mua}
\bibitem{Abe:2015mua}
  H.~Abe, T.~Kobayashi, H.~Otsuka and Y.~Takano,
  %``Realistic three-generation models from SO(32) heterotic string theory,''
  JHEP {\bf 1509} (2015) 056
  %doi:10.1007/JHEP09(2015)056
  [arXiv:1503.06770 [hep-th]].


%\cite{Abe:2016eyh}
\bibitem{Abe:2016eyh}
  H.~Abe, T.~Kobayashi, H.~Otsuka, Y.~Takano and T.~H.~Tatsuishi,
  %``Flavor structure in $SO(32)$ heterotic string theory,''
  Phys.\ Rev.\ D {\bf 94} (2016) no.12,  126020
  %doi:10.1103/PhysRevD.94.126020
  [arXiv:1605.00898 [hep-ph]].
  
  
  
%\cite{Blumenhagen:2005zg}
\bibitem{Blumenhagen:2005zg}
  R.~Blumenhagen, G.~Honecker and T.~Weigand,
  %``Non-Abelian brane worlds: The Heterotic string story,''
  JHEP {\bf 0510} (2005) 086
  [hep-th/0510049].
  
  
  
%\cite{Friedman:1997ih}
\bibitem{Friedman:1997ih}
  R.~Friedman, J.~W.~Morgan and E.~Witten,
  %``Vector bundles over elliptic fibrations,''
  alg-geom/9709029.
  
  
%\cite{Friedman:1997yq}
\bibitem{Friedman:1997yq}
  R.~Friedman, J.~Morgan and E.~Witten,
  %``Vector bundles and F theory,''
  Commun.\ Math.\ Phys.\  {\bf 187} (1997) 679
  [hep-th/9701162].



%\cite{Nibbelink:2015vha}
\bibitem{Nibbelink:2015vha}
  S.~Groot Nibbelink, O.~Loukas and F.~Ruehle,
  %``(MS)SM-like models on smooth Calabi-Yau manifolds from all three heterotic string theories,''
  Fortsch.\ Phys.\  {\bf 63} (2015) 609
  %doi:10.1002/prop.201500041
  [arXiv:1507.07559 [hep-th]].


%\cite{Ibanez:2001nd}
\bibitem{Ibanez:2001nd}
  L.~E.~Ibanez, F.~Marchesano and R.~Rabadan,
  %``Getting just the standard model at intersecting branes,''
  JHEP {\bf 0111} (2001) 002
  %doi:10.1088/1126-6708/2001/11/002
  [hep-th/0105155].

%\cite{Cvetic:2001tj}
\bibitem{Cvetic:2001tj}
  M.~Cvetic, G.~Shiu and A.~M.~Uranga,
  %``Three family supersymmetric standard - like models from intersecting brane worlds,''
  Phys.\ Rev.\ Lett.\  {\bf 87} (2001) 201801
  %doi:10.1103/PhysRevLett.87.201801
  [hep-th/0107143].
    
%\cite{Blumenhagen:2005mu}
\bibitem{Blumenhagen:2005mu}
  R.~Blumenhagen, M.~Cvetic, P.~Langacker and G.~Shiu,
  %``Toward realistic intersecting D-brane models,''
  Ann.\ Rev.\ Nucl.\ Part.\ Sci.\  {\bf 55} (2005) 71
  %doi:10.1146/annurev.nucl.55.090704.151541
  [hep-th/0502005].


%\cite{Honecker:2016gyz}
\bibitem{Honecker:2016gyz}
  G.~Honecker,
  %``From Type II string theory toward BSM/dark sector physics,''
  Int.\ J.\ Mod.\ Phys.\ A {\bf 31} (2016) no.34,  1630050
  %doi:10.1142/S0217751X16300507
  [arXiv:1610.00007 [hep-th]].
  

%\cite{Honecker:2012qr}
%\bibitem{Honecker:2012qr}
%  G.~Honecker, M.~Ripka and W.~Staessens,
  %``The Importance of Being Rigid: D6-Brane Model Building on $T^6/Z_2 x Z_6'$ with Discrete Torsion,''
%  Nucl.\ Phys.\ B {\bf 868} (2013) 156
  %doi:10.1016/j.nuclphysb.2012.11.011
%  [arXiv:1209.3010 [hep-th]].

  
  %\cite{Ecker:2014hma}
%\bibitem{Ecker:2014hma}
%  J.~Ecker, G.~Honecker and W.~Staessens,
  %``Rigour and rigidity: Systematics on particle physics D6-brane models on $Z_2 \times Z_6$,''
%  Fortsch.\ Phys.\  {\bf 62} (2014) 981
  %doi:10.1002/prop.201400066
%  [arXiv:1409.1236 [hep-th]].

  
  %\cite{Ecker:2015vea}
%\bibitem{Ecker:2015vea}
%  J.~Ecker, G.~Honecker and W.~Staessens,
  %``D6-brane model building on $\mathbb {Z}_2 \times \mathbb {Z}_6$: MSSM-like and left–right symmetric models,''
%  Nucl.\ Phys.\ B {\bf 901} (2015) 139
  %doi:10.1016/j.nuclphysb.2015.10.009
%  [arXiv:1509.00048 [hep-th]].

  

  %\cite{Candelas:1987kf}
\bibitem{Candelas:1987kf}
  P.~Candelas, A.~M.~Dale, C.~A.~Lutken and R.~Schimmrigk,
  %``Complete Intersection Calabi-Yau Manifolds,''
  Nucl.\ Phys.\ B {\bf 298} (1988) 493.
  %doi:10.1016/0550-3213(88)90352-5
  
  
  
%\cite{Candelas:1987du}
\bibitem{Candelas:1987du}
  P.~Candelas, C.~A.~Lutken and R.~Schimmrigk,
  %``Complete Intersection Calabi-yau Manifolds. 2. Three Generation Manifolds,''
  Nucl.\ Phys.\ B {\bf 306} (1988) 113.



    %\cite{Polchinsky}
\bibitem{Polchinsky}
  J.~Polchinski, String theory. Vol. 2: Superstring theory and beyond. Cambridge University Press, Cambridge,UK, (1998).   





  
%\cite{Freed:1986zx}
%\bibitem{Freed:1986zx}
%  D.~S.~Freed,
  %``Determinants, Torsion, and Strings,''
 % Commun.\ Math.\ Phys.\  {\bf 107} (1986) 483.


  %\cite{Witten:1998cd}
\bibitem{Witten:1998cd}
  E.~Witten,
  %``D-branes and K theory,''
  JHEP {\bf 9812} (1998) 019
  [hep-th/9810188].  
   


%\cite{Uranga:2000xp}
\bibitem{Uranga:2000xp}
  A.~M.~Uranga,
  %``D-brane probes, RR tadpole cancellation and K theory charge,''
  Nucl.\ Phys.\ B {\bf 598} (2001) 225
  [hep-th/0011048].
    

%\cite{Green:1987mn}
\bibitem{Green:1987mn}
  M.~B.~Green, J.~H.~Schwarz and E.~Witten,
  %``Superstring Theory. Vol. 2: Loop Amplitudes, Anomalies And Phenomenology,''
  Cambridge, Uk: Univ. Pr. ( 1987) 596 P. ( Cambridge Monographs On Mathematical Physics)
  

%\cite{Kobayashi}
\bibitem{Kobayashi}
  S.~Kobayashi,
  Publications of the Mathematical Society of Japan, Vol. 15, Iwanami Shoten, Tokyo Japan and Princeton University Press, Princeton U.S.A. (1987). 
  

%\cite{Green:1984bx}
\bibitem{Green:1984bx}
  M.~B.~Green, J.~H.~Schwarz and P.~C.~West,
  %``Anomaly Free Chiral Theories in Six-Dimensions,''
  Nucl.\ Phys.\ B {\bf 254} (1985) 327.
  
   
  %\cite{Ibanez:1986xy}
\bibitem{Ibanez:1986xy}
  L.~E.~Ibanez and H.~P.~Nilles,
  %``Low-Energy Remnants of Superstring Anomaly Cancellation Terms,''
  Phys.\ Lett.\  {\bf 169B} (1986) 354.  




  
  
%\cite{Braun:2010vc}
\bibitem{Braun:2010vc}
  V.~Braun,
  %``On Free Quotients of Complete Intersection Calabi-Yau Manifolds,''
  JHEP {\bf 1104} (2011) 005
%  doi:10.1007/JHEP04(2011)005
  [arXiv:1003.3235 [hep-th]].


  %\cite{Witten:1995gx}
\bibitem{Witten:1995gx}
  E.~Witten,
  %``Small instantons in string theory,''
  Nucl.\ Phys.\ B {\bf 460} (1996) 541
  [hep-th/9511030].

  

%\cite{Ambroso:2009jd}
\bibitem{Ambroso:2009jd}
  M.~Ambroso and B.~Ovrut,
  %``The B-L/Electroweak Hierarchy in Heterotic String and M-Theory,''
  JHEP {\bf 0910} (2009) 011
  %doi:10.1088/1126-6708/2009/10/011
  [arXiv:0904.4509 [hep-th]].
  

%\cite{Ovrut:2012wg}
\bibitem{Ovrut:2012wg}
  B.~A.~Ovrut, A.~Purves and S.~Spinner,
  %``Wilson Lines and a Canonical Basis of SU(4) Heterotic Standard Models,''
  JHEP {\bf 1211} (2012) 026
  %doi:10.1007/JHEP11(2012)026
  [arXiv:1203.1325 [hep-th]].
    
  
%\cite{Candelas:1987se}
\bibitem{Candelas:1987se}
  P.~Candelas,
  %``Yukawa Couplings Between (2,1) Forms,''
  Nucl.\ Phys.\ B {\bf 298} (1988) 458.



  

%\cite{Anderson:2009ge}
\bibitem{Anderson:2009ge}
  L.~B.~Anderson, J.~Gray, D.~Grayson, Y.~H.~He and A.~Lukas,
  %``Yukawa Couplings in Heterotic Compactification,''
  Commun.\ Math.\ Phys.\  {\bf 297} (2010) 95
  %doi:10.1007/s00220-010-1033-8
  [arXiv:0904.2186 [hep-th]].
  

%\cite{Blesneag:2015pvz}
\bibitem{Blesneag:2015pvz}
  S.~Blesneag, E.~I.~Buchbinder, P.~Candelas and A.~Lukas,
  %``Holomorphic Yukawa Couplings in Heterotic String Theory,''
  JHEP {\bf 1601} (2016) 152
  %doi:10.1007/JHEP01(2016)152
  [arXiv:1512.05322 [hep-th]].

%\cite{Blesneag:2016yag}
\bibitem{Blesneag:2016yag}
  S.~Blesneag, E.~I.~Buchbinder and A.~Lukas,
  %``Holomorphic Yukawa Couplings for Complete Intersection Calabi-Yau Manifolds,''
  JHEP {\bf 1701} (2017) 119
  %doi:10.1007/JHEP01(2017)119
  [arXiv:1607.03461 [hep-th]].
  

%\cite{Anderson:2014hia}
\bibitem{Anderson:2014hia}
  L.~B.~Anderson, A.~Constantin, S.~J.~Lee and A.~Lukas,
  %``Hypercharge Flux in Heterotic Compactifications,''
  Phys.\ Rev.\ D {\bf 91} (2015) no.4,  046008
  doi:10.1103/PhysRevD.91.046008
  [arXiv:1411.0034 [hep-th]].
  

%\cite{Abe:2015xua}
\bibitem{Abe:2015xua}
  H.~Abe, T.~Kobayashi, H.~Otsuka, Y.~Takano and T.~H.~Tatsuishi,
  %``Gauge coupling unification in SO(32) heterotic string theory with magnetic fluxes,''
  PTEP {\bf 2016} (2016) no.5,  053B01
  %doi:10.1093/ptep/ptw038
  [arXiv:1507.04127 [hep-ph]].






  

  



\end{thebibliography}
\end{document}